%% file: template.tex
\documentclass{article}

\usepackage{arxiv}

\usepackage[utf8]{inputenc} % allow utf-8 input
\usepackage[T1]{fontenc}    % use 8-bit T1 fonts
\usepackage{hyperref}       % hyperlinks
\usepackage{url}            % simple URL typesetting
\usepackage{booktabs}       % professional-quality tables
\usepackage{amsfonts}       % blackboard math symbols
\usepackage{nicefrac}       % compact symbols for 1/2, etc.
\usepackage{microtype}      % microtypography
\usepackage{lipsum}
\usepackage{graphicx}
\graphicspath{ {./images/} }
\usepackage{comment}
\usepackage{listings}
\usepackage[linesnumbered,ruled]{algorithm2e}
\usepackage{algorithmic}
\usepackage{multirow}
\usepackage{subfigure}
\graphicspath{{figure/}}
\usepackage{wrapfig}
\usepackage{siunitx}
\usepackage{pifont}
\usepackage{enumitem}
\usepackage{amsmath}

\title{gECC: A GPU-based high-throughput framework for Elliptic Curve Cryptography}

\author{
 Qian Xiong \\
  Huazhong University \\
  of Science and Technology, \\
  Wuhan, China\\
  \texttt{xiongqian@hust.edu.cn} \\
   \And
 Weiliang Ma \\
  Huazhong University \\
  of Science and Technology, \\
  Wuhan, China\\
  \texttt{weiliangma@hust.edu.cn} \\
  \And
 Xuanhua Shi \\
  Huazhong University \\
  of Science and Technology, \\
  Wuhan, China\\
  \texttt{xhshi@hust.edu.cn} \\
  \And
 Yongruan Zhou \\
  University of Copenhagen \\
  Copenhagen, Denmark\\
  \texttt{zhou@di.ku.dk} \\
  \And
 Hai Jin \\
  Huazhong University \\
  of Science and Technology, \\
  Wuhan, China\\
  \texttt{hjin@hust.edu.cn} \\
  \And
 Kaiyi Huang \\
  Huazhong University \\
  of Science and Technology, \\
  Wuhan, China\\
  \texttt{kyhuang@hust.edu.cn} \\
  \And
 Haozhou Wang \\
  Huazhong University \\
  of Science and Technology, \\
  Wuhan, China\\
  \texttt{whz\_tec003@hust.edu.cn} \\
  \And
 Zhengru Wang \\
  Nvidia \\
  Shanghai, China\\
  \texttt{zhengruw@nvidia.com} \\
}

\begin{document}
\input{macro}
\maketitle

\input{abstract}


\input{introduction}

\input{background}

\input{alg_layer}

\input{modular_design}

\input{evaluation}

\input{conclusion}

\bibliographystyle{unsrt}  
\bibliography{references}  %%% Remove comment to use the external .bib file (using bibtex).
%%% and comment out the ``thebibliography'' section.

%%% Comment out this section when you \bibliography{references} is enabled.
% \begin{thebibliography}{1}

% \bibitem{kour2014real}
% George Kour and Raid Saabne.
% \newblock Real-time segmentation of on-line handwritten arabic script.
% \newblock In {\em Frontiers in Handwriting Recognition (ICFHR), 2014 14th
%   International Conference on}, pages 417--422. IEEE, 2014.

% \bibitem{kour2014fast}
% George Kour and Raid Saabne.
% \newblock Fast classification of handwritten on-line arabic characters.
% \newblock In {\em Soft Computing and Pattern Recognition (SoCPaR), 2014 6th
%   International Conference of}, pages 312--318. IEEE, 2014.

% \bibitem{hadash2018estimate}
% Guy Hadash, Einat Kermany, Boaz Carmeli, Ofer Lavi, George Kour, and Alon
%   Jacovi.
% \newblock Estimate and replace: A novel approach to integrating deep neural
%   networks with existing applications.
% \newblock {\em arXiv preprint arXiv:1804.09028}, 2018.

% \end{thebibliography}

\end{document}

%% file: macro.tex
%!TEX root = main.tex
% only foreign words should be italicized
% \newcommand{\eg}{\textit{e.g.}}
% \newcommand{\ie}{\textit{i.e.}}
\newcommand{\etal}{\textit{et al.}}
\newcommand{\etc}{\textit{etc.}}
\newcommand{\adhoc}{\textit{ad hoc}}
\newcommand{\viz}{\textit{viz.}}

% general
\newcommand{\XXX}{{\bf XXX}}
\newcommand{\sref}[1]{\S\ref{#1}}
\newcommand{\oursys}{gECC\xspace} % or cuECCL
\newcommand{\sys}{gECC\xspace}
\newcommand{\msm}{multi-scalar multiplication}
\newcommand{\sm}{scalar multiplication}

\newcommand{\ecc}{ECC\xspace}
\newcommand{\ec}{EC\xspace}
\newcommand{\rsa}{RSA}
\newcommand{\nvcc}{NVCC}

% baseline
\newcommand{\libsnark}{libsnark}
\newcommand{\othergpu}{MINA}
\newcommand{\bellman}{bellman}
\newcommand{\bellperson}{bellperson\xspace}

% axes
\newcommand{\xaxis}{$x$-axis}
\newcommand{\yaxis}{$y$-axis}

% table
\newcommand{\tabincell}[2]{\begin{tabular}{@{}#1@{}}#2\end{tabular}}

% evaluation data
\newcommand{\maxbkGPU}{48.1$\times$\xspace}
\newcommand{\avgbkGPU}{33.6$\times$\xspace}
\newcommand{\maxzcashGPU}{17.6$\times$\xspace}
\newcommand{\avgzcashCPU}{24.7$\times$\xspace}
\newcommand{\avgzcashGPU}{13.2$\times$\xspace}
\newcommand{\maxNTT}{318$\times$\xspace}
\newcommand{\maxMSM}{29$\times$\xspace}
\newcommand{\avgNTTCPU}{180$\times$\xspace}
\newcommand{\avgNTTGPU}{5.8$\times$\xspace}
\newcommand{\maxNTTGPU}{10.3$\times$\xspace}
% compare with mina scheme
% \newcommand{\avgMSM}{19x}
% compare with bellperson
\newcommand{\avgMSMGPU}{9.1$\times$\xspace}
\newcommand{\maxMSMGPU}{17.9$\times$\xspace}
% units
\newcommand{\KB}{~KB}
\newcommand{\MB}{~MB}
\newcommand{\GB}{~GB}
\newcommand{\MBs}{~MB/s}
\newcommand{\mus}{$\mu s$}

% commands and calls
\newcommand{\unix}{{\sc Unix}}
\newcommand{\fsck}{\texttt{fsck}}
\newcommand{\fsync}{\texttt{fsync}}
\newcommand{\myread}{\texttt{read}}
\newcommand{\mywrite}{\texttt{write}}
\newcommand{\mysync}{\texttt{sync}}
\newcommand{\txwrite}{\texttt{txwrite}}

% others
\newcommand{\Term}[1]{{\em #1}}
\newcommand{\Phrase}[1]{{\it #1}}
\newcommand{\File}[1]{{\tt #1}}
\newcommand{\Keyword}[1]{{\small \tt #1}}
\newcommand{\Symbol}[1]{{\tt #1}}
\newcommand{\CBox}[1]{{\parbox{\columnwidth}{#1}}}
\newcommand{\CaptionKeyword}[1]{{\footnotesize \sf #1}}
\newcommand{\yes}{{$\surd$}}

% Acknowledgments
% \newcommand\acksname{Acknowledgments}
% \specialcomment{acks}{%
%   \begingroup
%   \section*{\acksname}
%   \phantomsection\addcontentsline{toc}{section}{\acksname}
% }{%
%   \endgroup
% }

% to add space less than a line
\newcommand{\halfline}{\vspace{6 pt}}
\newcommand{\quartline}{\vspace{3 pt}}
\newcommand{\smallline}{\vspace{2 pt}}
\newcommand{\tinyline}{\vspace{1 pt}}

% bullets and labels
\newcommand{\mybullet}[1]{
\vspace{0.02in}
\noindent$\bullet$
%\hspace{0.1cm}
{\bf #1}}

\newcommand{\mybulletnobold}[1]{
\vspace{0.02in}
\noindent$\bullet$
%\hspace{0.1cm}
{#1}}

\newcommand{\mylabel}[2]{
\vspace{0.02in}
\noindent {\bf #1} {#2}
\hspace{0.01cm}}

\newcommand{\mylabelemph}[1]{
\vspace{0.02in}
\noindent {\em #1}
%\hspace{0.01cm}
}

\newcommand{\mylabelbold}[1]{
\vspace{0.02in}
\noindent {\bf #1}
%\hspace{0.01cm}
}

% captions
\newcommand{\beforecaption}{\vspace{-.15cm}\begin{spacing}{0.85}}
\newcommand{\aftercaption}{\vspace{-.45cm}\end{spacing}}
\newcommand{\mycaption}[3]{{
\beforecaption
\caption{\label{#1} \footnotesize{\textbf{#2}} {\em #3}}
\aftercaption}}

% some more captions
\newcommand{\beforecaptiontwo}{\vspace{0.00cm}\begin{spacing}{0.95}}
\newcommand{\aftercaptiontwo}{\vspace{0.00cm}\end{spacing}}
\newcommand{\mycaptiontwo}[3]{{
\beforecaptiontwo
\caption{\label{#1} {\bf #2} {\em #3}}
\aftercaptiontwo}}

% system
\newcommand{\System}[1]{{\footnotesize {\sf #1}}}

% define this for small code snippets
\newcommand{\code}[1]{{\texttt{#1}}}

% less white space for itemize
\newenvironment{tightitemize}{%
\begin{list}{$\bullet$}{%
\setlength{\topsep}{3pt}%
\setlength{\parskip}{0pt}%
\setlength{\parsep}{0pt}%
\renewcommand{\makelabel}[1]{\textbf{##1} }
\setlength{\labelwidth}{0pt}%
\setlength{\leftmargin}{5pt}%
\setlength{\labelsep}{0pt}%
\setlength{\listparindent}{0pt}%
\setlength{\columnsep}{0.25in}
}}%
{\end{list}}

\newcommand{\VP}[1]{{\bf #1}}
\newcommand{\TODO}[1]{{\color{red} {\bf #1}}}
\newcommand{\inlcode}[1]{{\small{\texttt{#1}}}}
\definecolor{darkgray}{rgb}{0.16,0.16,0.16}

%% XM's defs
\newcommand{\cb}{\textcolor{blue}}
\newcommand{\corange}{\textcolor{orange}}
\newcommand{\cred}{\textcolor{red}}
\definecolor{darkgreen}{rgb}{0.078,0.667,0.016}
\newcommand{\cg}{\textcolor{darkgreen}}
\newcommand{\cp}{\textcolor{purple}}

\newcommand{\mwl}[1]{\cred{[~mwl~:~#1~]}}
\newcommand{\tofill}{\cred{@@@}\xspace}
\newcommand{\tocite}{\cred{\cite{}}\xspace}
\newcommand{\toref}{\cred{\ref{}}\xspace}
\newcommand{\para}[1]{\noindent \textbf{#1 }}
\newcommand{\outline}[1]{}
\newcommand{\Eg}{\textit{E.g.}\xspace}
\newcommand{\eg}{\textit{e.g.}\xspace}
\newcommand{\ie}{\textit{i.e.}\xspace}

%% responds's def
\newcommand{\xq}[1]{\cg{(#1 --XQ)}\xspace}

\renewcommand{\algorithmicrequire}{\textbf{Input:}}
\renewcommand{\algorithmicensure}{\textbf{Output:}}

%\lstdefinestyle{customc}{
%	belowcaptionskip=1\baselineskip,
%	breaklines=true,
%	frame=L,
%	xleftmargin=\parindent,
%	language=python,
%	showstringspaces=false,
%	basicstyle=\footnotesize\ttfamily,
%	keywordstyle=\bfseries\color{green!40!black},
%	commentstyle=\itshape\color{purple!40!black},
%	identifierstyle=\color{blue},
%	stringstyle=\color{orange},
%}
%\lstset{escapechar=@,style=customc}

% Define Colors
% Define Language
\lstdefinelanguage{myLang}
{
	% list of keywords
	morekeywords={
		import,
		if,
		while,
		for,
		void,
		return,
		val,
		new
	},
	sensitive=false, % keywords are not case-sensitive
	morecomment=[l]{//}, % l is for line comment
	morecomment=[s]{/*}{*/}, % s is for start and end delimiter
	morestring=[b]" % defines that strings are enclosed in double quotes
}

\definecolor{customPurple}{RGB}{127,0,127}
\definecolor{customGreen}{RGB}{0,100,0}
\definecolor{customBlue}{RGB}{0,0.0,255}

% Set Language
\lstset{
	language={myLang},
	basicstyle=\fontsize{8}{8}\selectfont\ttfamily, % Global Code Style
%	\bfseries \bfseries\ttfamily\scriptsize\footnotesize
%	alsoletter={0,1,2,3,4,5,6,7,8,9},
%	morekeywords=[2]{1,2,3,1000},
%	keywordstyle=[2]\color{orange},
%	literate={1000}{{{\color{orange}1000}}}4,
	captionpos=b, % Position of the Caption (t for top, b for bottom)
	extendedchars=true, % Allows 256 instead of 128 ASCII characters
	tabsize=4, % number of spaces indented when discovering a tab 
	columns=fixed, % make all characters equal width
	keepspaces=true, % does not ignore spaces to fit width, convert tabs to spaces
	showstringspaces=false, % lets spaces in strings appear as real spaces
	breaklines=true, % wrap lines if they don't fit
%	frame=none, % draw a frame at the top, right, left and bottom of the listing
%	frameround=tttt, % make the frame round at all four corners
%	framesep=4pt, % quarter circle size of the round corners
%	numbers=left, % show line numbers at the left
%	numberstyle=\tiny\ttfamily, % style of the line numbers
	commentstyle=\color{customGreen}, % style of comments
	keywordstyle=\color{customBlue}, % style of keywords
	stringstyle=\color{customPurple}, % style of strings
}

\newcommand\bverb[1][]{\begingroup\ifx\relax#1\relax
    \def\tmpA{\allowbreak}\let\tmpB=\empty 
      \setlength\spaceskip{1.5ex plus 1ex minus .5ex}% SETTING \xspaceskip CAN BE DONE
    \else\def\tmpA{}\let\tmpB=#1 
      \setlength\spaceskip{1.5ex plus 1ex minus .5ex}\fi% SETTING \xspaceskip CAN BE DONE
  \catcode`\%=12 \bverbaux}

\newcommand\bverbaux[2][ ]{%
   \ttfamily\tokencycle
   {\tctestifx{\tmpB##1}{\addcytoks{\allowbreak}}{%
    \addcytoks[1]{\tmpA}\addcytoks[1]{\string##1}%
    \addcytoks{\nobreak\hspace{0pt plus 5pt minus .5pt}}}}
   {\addcytoks[1]{\string{}\processtoks{##1}\addcytoks[1]{\string}}}
   {\tctestifx{\tmpB##1}{\addcytoks{\allowbreak}}
     {\tctestifnum{\cstest{##1}=11} 
      {\tcpeek\zzz\tctestifcatnx\zzz A{\tcpush{\space}}{}}{}%
      \tcpush{\string##1}}}
   {\addcytoks{\allowbreak#1}}#2\endtokencycle\endgroup}
\def\cstest#1{\expandafter\cstestaux\string#1.\relax}
\def\cstestaux#1#2#3\relax{\the\catcode`#2 }

%% file: abstract.tex
\begin{abstract}
Elliptic Curve Cryptography (ECC) is an encryption method that provides security comparable to traditional techniques like Rivest–Shamir–Adleman (RSA) but with lower computational complexity and smaller key sizes, making it a competitive option for applications such as blockchain, secure multi-party computation, and database security. However, the throughput of \ecc{} is still hindered by the significant performance overhead associated with elliptic curve (\ec{}) operations, which can affect their efficiency in real-world scenarios. This paper presents \oursys{}, a versatile framework for \ecc{} optimized for GPU architectures, specifically engineered to achieve high-throughput performance in EC operations. To maximize throughput, \oursys{} incorporates batch-based execution of \ec{} operations and microarchitecture-level optimization of modular arithmetic. It employs Montgomery’s trick~\cite{montgomery1987speeding} to enable batch \ec{} computation and incorporates novel computation parallelization and memory management techniques to maximize the computation parallelism and minimize the access overhead of GPU global memory. Furthermore, we analyze the primary bottleneck in modular multiplication by investigating how the user codes of modular multiplication are compiled into hardware instructions and what these instructions' issuance rates are. We identify that the efficiency of modular multiplication is highly dependent on the number of Integer Multiply-Add (IMAD) instructions. To eliminate this bottleneck, we propose novel techniques to minimize the number of IMAD instructions by leveraging predicate registers to pass the carry information and using addition and subtraction instructions (IADD3) to replace IMAD instructions. Our experimental results show that, for ECDSA and ECDH, the two commonly used \ecc{} algorithms, \oursys{} can achieve performance improvements of 5.56 $\times$ and 4.94 $\times$, respectively, compared to the state-of-the-art GPU-based system. In a real-world blockchain application, we can achieve performance improvements of 1.56 $\times$, compared to the state-of-the-art CPU-based system.
\oursys{} is completely and freely available at \url{https://github.com/CGCL-codes/gECC}.
\end{abstract}

%% file: introduction.tex
\section{Introduction}

Elliptic Curve Cryptography (\ecc{})~\cite{miller1985use, koblitz1987elliptic} is a method for public-key encryption by using elliptic curves. It offers security that is on par with conventional public-key cryptographic systems like Rivest–Shamir–Adleman (\rsa{}) but with a smaller key size and lower computational complexity. 
% This makes \ecc{} a more efficient choice for resource-limited settings such as Internet of Things devices~\cite{suarez2018practical}. 
Recently, \ecc{} gained increased attention due to its efficient privacy protection and fewer interactions in verifiable databases~\cite{yue2022glassdb, ge2022hybrid, loghin2022anatomy, yang2020ledgerdb, zhang2020pvldb}. 

ECC serves as a powerful encryption tool widely used in areas such as data encryption, digital signatures, blockchain, secure transmission, and secure multi-party computation. The two most prevalent public-key algorithms based on \ecc{} are Elliptic Curve Diffie-Hellman (ECDH)~\cite{blake2006elliptic} for data encryption and Elliptic Curve Digital Signature Algorithm (ECDSA)~\cite{johnson2001elliptic} for digital signatures.
Transport Layer Security (TLS), the preferred protocol for securing 5G communications, incorporates ECDH in its handshake process~\cite{blake2006elliptic}. ECDH also plays a crucial role in sensitive data storage~\cite{8840125, suthanthiramani2021secured} and data sharing, allowing basic database query instructions to be executed without any data information leakage. This includes Private Set Intersection (PSI)~\cite{pinkas2019spot, psichina, lu2023efficient} and Private Intersection Sum~\cite{li2024pis}.
ECDSA is widely employed in blockchain systems to safeguard data integrity and transaction accountability~\cite{wang2019cryptographic}. Currently, numerous blockchain-database (referred to as verifiable databases)~\cite{yue2022glassdb, ge2022hybrid, loghin2022anatomy, yang2020ledgerdb, zhang2020pvldb}, which combine the properties of blockchains with the ones of classic Database Management Systems (DBMS), protect data history against malicious tampering. Additionally, researchers developed verifiable SQL~\cite{zhang2015integridb, pang2009scalable}, protecting the integrity of user data and query execution on untrusted database providers. As an up-and-coming field approaching commercial applications, cloud service providers, such as Amazon~\cite{Amazon2019} and Microsoft~\cite{antonopoulos2021sql}, provide services that can maintain an append-only and cryptographically verifiable log 
of data operations. %, which  %similar to that of certificate transparency to record data operations that are
%then applied to another backend database.

To enhance the efficiency of \ecc{}, researchers have dedicated significant efforts to designing specialized curve forms that reduce the computational overhead associated with modular arithmetic in elliptic curve operations. For instance, blockchain systems, such as Bitcoin~\cite{nakamoto2008bitcoin}, Ethereum~\cite{wood2014ethereum}, Zcash~\cite{sasson2014zerocash}, and Hyperledger Fabric~\cite{androulaki2018hyperledger}, use the secp256k1 curve~\cite{qu1999sec} and the P-256 curve endorsed by the National Institute of Standards and Technology (NIST)~\cite{faqs2024federal} for ECDSA. Differently, China keeps the SM2 curve as the standard for electronic authentication systems, key management, and e-commerce applications~\cite{yang2014provably, state2016public}. 

Despite such progress, \ecc{} remains a bottleneck in the performance of these throughput-sensitive applications. On mainstream server environments, a single fundamental \ec{} operation typically takes more than 6 milliseconds to execute. A PSI computation to identify the intersection between two datasets containing millions of items has to perform tens of millions of \ec{} operations, which could take around 632 seconds to complete~\cite{pinkas2019spot}. Although researchers develop blockchain-database systems with ASIC~\cite{vdbs23Baldur} to accelerate ECDSA to improve transaction throughput, the improvement compared to a CPU-based system is limited, with a maximal 12\% improvement.

%They overlook the potential to reduce the complexity of \ec{} operations by \cb{exploiting Montgomery’s trick to batch \ec{} operations}, which can further enhance the throughput of \ec{} operations; they involve costly discontinuous accesses in GPU global memory and do not utilize multi-level caches to minimize memory access overhead on new GPU architecture; their underlying modular arithmetic employs numerous high-latency instructions and lack specific optimizations for prime moduli.

\begin{wrapfigure}{ r }{ 0.5 \textwidth } 
    % \vspace{-0.5cm}
    \centering
    \setlength{\abovecaptionskip}{0.1cm}
    \includegraphics[width=0.5\textwidth]{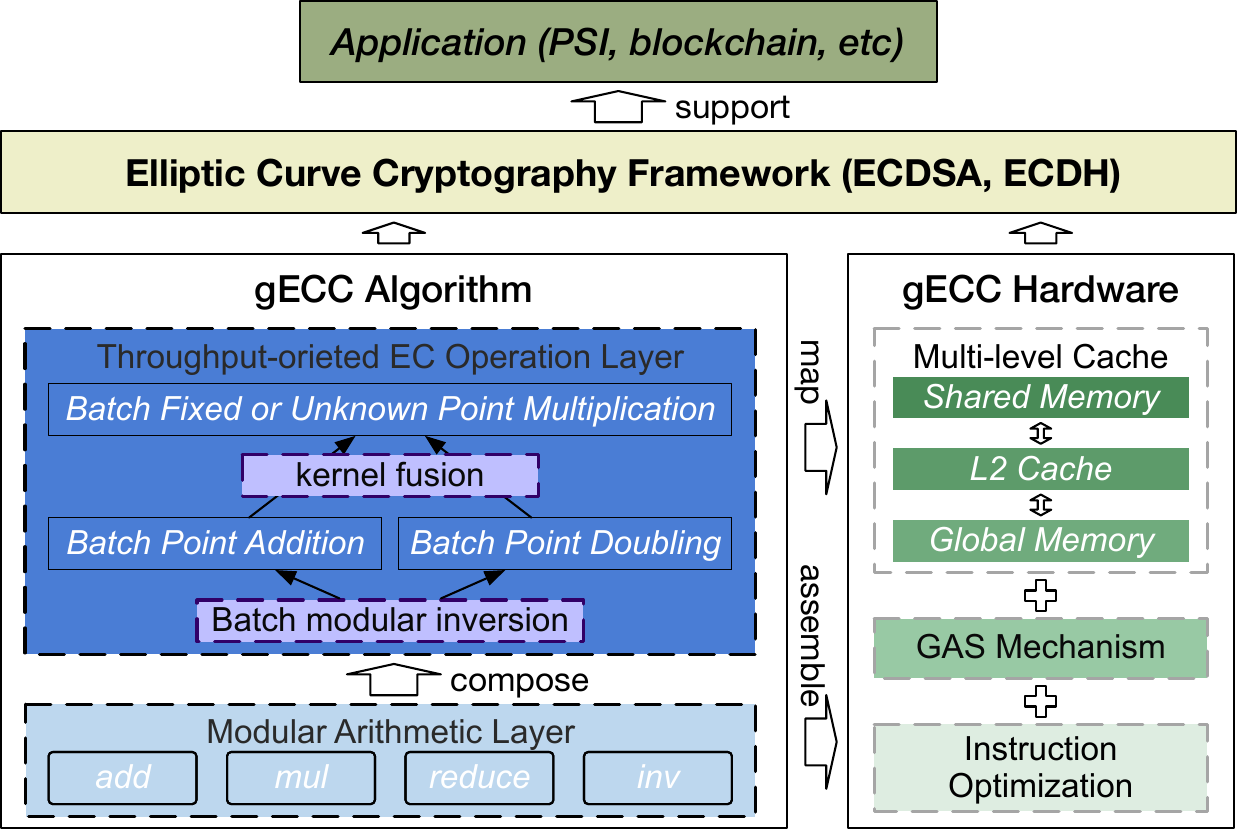}
    \caption{Overview of \sys Framework.} 
    \label{fig:workflow of library}
    \vspace{-0.25cm}
\end{wrapfigure}
There are recent efforts~\cite{gao2020dpf, gao2021dpf, feng2022accelerating} in employing GPU to minimize the latency of individual \ec{} operations. However, achieving the high-throughput requirements of emerging big data applications remains a challenge. To close the gap, we present a high-throughput GPU-based \ecc{} framework, which is holistically optimized to maximize the data processing throughput. The optimizations consist of the following three major aspects. 

First, we employ Montgomery’s trick~\cite{montgomery1987speeding} to batch \ec{} operations to enhance the overall throughput. Existing GPU-based \ecc{} solutions solely offer interfaces for individual EC operations, such as single-point addition and single-point multiplication, and therefore they cannot bach \ec{} operations. We reconstruct the entire \ecc{} framework, called
\oursys{}, consisting of batch \ec{} operations, including point multiplication, point addition, and point doubling.
However, the modular inversion operation in the algorithm incurs significant overhead due to poor parallelism, thus reducing performance.
To address the issue, we borrow the wisdom of parallel graph processing systems and incorporate the Gather-Apply-Scatter (GAS) mechanism~\cite{joseph2012powergraph} to reduce the overhead of the modular inversion operation. To the best of our knowledge, \oursys{} is the first system to implement batch PMUL operations on a GPU using Montgomery's trick.
%Additionally, we employ locality-aware kernel fusion optimization and multi-level cache management to minimize memory access overhead.
%but the modular inversion operation in the algorithm incurs significant overhead due to poor parallelism, thus reducing performance.}

Second, we employ data-locality-aware kernel fusion optimization and design multi-level cache management to minimize the memory access overhead incurred by the frequent data access for point data and the large intermediate results caused by batching \ec{} operations with Montgomery's Trick.
Each EC operation computation inherently requires access to the point data and the intermediate results twice.
These data need more than hundreds of megabytes of memory space, which far exceeds the limited GPU's shared memory size.
% the large number of intermediate results generated by batch \ec{} operations with Montgomery's Trick. 
% Each EC operation computation inherently requires access to the intermediate result that is 5 times the size of the input. With Montgomery's Trick, an additional 4 times the size of the input of intermediate results is needed. 
For example, when $2^{20}$ EC point additions are performed simultaneously, the total size of data to be temporarily stored is \SI{96}{MB} and the inputs are \SI{128}{MB}. Our techniques optimize data access to minimize the pressure on the registers and GPU's shared memory.

Third, all the \ec{}'s arithmetic operations, including addition, subtraction, and multiplication, are based on modular arithmetic on the finite field with at least 256-bit. Among these, the most time-consuming modular multiplication operation is also the most frequently performed operation in all kinds of \ec{} operations. Previous studies~\cite{gao2020dpf, gao2021dpf, feng2022accelerating, cgbn, emmart2018faster} fall short in evaluating the performance of modular multiplication from the perspective of instructions' issuance rates at the microarchitectural level. 
%, but rather only from the application level. 
%These outcomes lead to the introduction of high-latency instructions, which further exacerbate performance issues. 
Furthermore, there is a noticeable absence of arithmetic optimization for specific prime moduli in the existing work~\cite{feng2022accelerating, emmart2018faster}. We identify that the efficiency of modular multiplication is highly dependent on the
number of Integer Multiply-Add (IMAD) instructions. To eliminate this bottleneck, we propose novel techniques to minimize the
number of IMAD instructions by leveraging predicate registers to pass the carry information. In addition, we develop an arithmetic optimization for the SM2 curve, using addition and subtraction
instructions (IADD3) to replace the expensive IMAD instructions.

We implement \oursys{} using CUDA and conduct evaluations on the Nvidia A100 GPU to assess its performance. Our comparative analysis includes \oursys{} and the leading GPU-based \ecc{} system, RapidEC~\cite{feng2022accelerating}, highlighting the advancements and improvements offered by our framework. %As detailed in Section~\ref{sec:eval}, 
On standard cryptographic benchmarks for two commonly used \ecc{} algorithms, our results demonstrate that \oursys{} achieves a speedup of $4.18 \times$ for signature generation and $5.56 \times$ for signature verification in ECDSA. If we measure the two key layers in our framework separately, \oursys{} achieves up to $4.04 \times$ and $4.94 \times$ speedup for fixed-point multiplication and unknown point multiplication of the EC operator layer, and $1.72 \times$ speedup in the modular multiplication for SM2 curve in the modular arithmetic layer. In real-world blockchain system, we can achieve 1.56$\times$ performance improvements, compared to the state-of-the-art CPU-based systems.
\oursys{} is completely and freely available at \url{https://github.com/CGCL-codes/gECC}.

%% file: background.tex
\section{Background and Related Work}

% \begin{table}[ht]
%     \centering
%     \caption{List of Symbols}
%     \label{tab:symbols}    
%     \begin{tabular}{c|c}
%         \hline
%         \textbf{Symbol} & \textbf{Description} \\ \hline
%         $q$ & A large prime moduli \\ 
%         $F_q$ & A finite field, integers range from $[0, q-1]$ \\
%         $D$ & integer $x$ base $D$ representation, $x  =\sum_{i=0}^{m-1}x_i D^i$ \\
%         \verb|X[m]| & integer $x$ storage structure, \verb|X[m]| $=\{x_0 \dots x_{m-1}\}$ \\
%         \hline
%     \end{tabular}
% \end{table}

\subsection{Elliptic Curve Cryptography}
\ecc{} is a form of public-key cryptography that relies on the algebraic properties of elliptic curves over finite fields. It is widely employed in various cryptographic algorithms and protocols, including key exchange, digital signatures, and others. \ecc{} is also utilized in several standard algorithms, such as ECDSA and ECDH. One of the key features of \ecc{} is its ability to offer the same level of security as \rsa{} encryption with smaller key sizes, making it particularly appealing for resource-constrained environments.
% Consequently, \ecc{} has applications in diverse scenarios, ranging from secure communications to digital currency and database applications.

The SM2 elliptic curve~\cite{state2016public} utilized in ECDSA applications is defined by Eqn.~\ref{form:ec-equation} 
% and is classified as \emph{Short Weierstrass curve}
.
The parameters $a$ and $b$ are elements in a finite field $F_q$, and they establish the specific definition of the curve. A pair $(x, y)$, where $(x, y \in F_q)$, is considered a point on the curve if it satisfies the aforementioned Eqn.~\ref{form:ec-equation}.
\begin{equation}
\label{form:ec-equation}
    \begin{aligned}
        y^2 &= x^3 + ax + b,\ Short\ Weierstrass\ curve 
    \end{aligned}
    % \vspace{-0.6cm}
\end{equation}
\subsubsection{Elliptic Curve Point Operations}
The fundamental \ec{} operations are point addition (PADD) and point doubling (PDBL), which are essential building blocks for more complex \ec{} operations, such as point multiplication (PMUL), forming the basis for various cryptographic protocols in \ecc{}.

\textbf{PADD operation.} 
% As indicated in formula~\ref{form:point-add-1}, 
This operation involves adding two points, P$(x_p, y_p)$ and T$(x_t, y_t)$, on an elliptic curve to obtain the resulting point R$(x_r, y_r) = (x_p, y_p) + (x_t, x_t)$. The addition is performed using the following two equations: 1) Eqn.~\ref{form:point-add-1} is to calculate $\lambda$. 2) Eqn.~\ref{form:point-add-2} is to calculate the values of X and Y coordinates of the result point $R$, respectively.
\begin{equation}
\label{form:point-add-1}
    \begin{aligned}
        \lambda &= \frac{y_p - y_t}{x_p - x_t}
        % \lambda &= \frac{y_p - y_t}{x_p - x_t} &\quad
        % x_r &= \lambda^2 - x_p - x_t &\quad
        % y_r &= \lambda(x_p-x_r) - y_p
    \end{aligned}
\end{equation}
\begin{equation}
\label{form:point-add-2}
    \begin{aligned}
        % \lambda &= \frac{y_p - y_t}{x_p - x_t} &\quad
        x_r &= \lambda^2 - x_p - x_t &\quad
        y_r &= \lambda(x_p-x_r) - y_p
    \end{aligned}
\end{equation}
% \begin{equation}
% \label{form:point-add-3}
%     \begin{aligned}
%         % \lambda &= \frac{y_p - y_t}{x_p - x_t} &\quad
%         % x_r &= \lambda^2 - x_p - x_t &\quad
%         y_r &= \lambda(x_p-x_r) - y_p
%     \end{aligned}
% \end{equation}

\textbf{PDBL operation.} This operation is used when the points P and T are identical. In this case, $\lambda$ is calculated differently, by $\frac{3{x_p}^2 + a}{2y_p}$. Here, $a$ is the parameter of the elliptical curve Eqn.~\ref{form:ec-equation}, and the rest of the PDBL operation follows the same steps as the PADD operation using this new value of $\lambda$.

% \begin{wraptable}{r}{0.8\linewidth}
\begin{table}[h]
    % \vspace{-0.5cm}
    \centering
    \caption{Performance analysis of \ec{} operations on Short Weierstrass curves in different coordinate systems}
    \label{tab:coord-select}
    \begin{tabular}{ccc}
        \toprule %[2pt] 
        \textbf{operation} & \textbf{affine coordinate} & \textbf{Jacobian coordinate}\\ \hline
        % \midrule %[2pt]
        PADD & 1 $modinv$ + 3 $modmul$ + 6 $modadd$ & 11 $modmul$ + 6 $modadd$ \\ \hline
        % \midrule %[2pt]
        PDBL & 1 $modinv$ + 4 $modmul$ + 5 $modadd$ & 10 $modmul$ + 10 $modadd$ \\ 
        \bottomrule %[2pt] 
    \end{tabular}
    \vspace{-0.25cm}
% \end{wraptable}
\end{table}
Performing \ec{} operations on points represented in the original coordinate system (called \textit{affine coordinate}), such as Eqn.~\ref{form:point-add-1}, \ref{form:point-add-2}, results in lower efficiency. This is mainly due to the time-consuming nature of modular inversion operations when computing $\frac{1}{x_p - x_t}$, significantly affecting overall performance. 
Various types of elliptic curve coordinate systems have been introduced to enhance the computational efficiency of single \ec{} operations. For instance, the \textit{Jacobian coordinate system}, used in~\cite{feng2022accelerating, openssl2024, pan2017efficient}, represents points using the triplets $(X,Y,Z)$, where $x = X/Z^2$ and $y = Y/Z^3$~\cite{hankerson2006guide}. This coordinate system eliminates the need for costly modular inversion and improves the performance of single PADD and PDBL operations.
% , as demonstrated in Table~\ref{tab:coord-select}. 
Table~\ref{tab:coord-select} lists the number of modular multiplication ($modmul$), modular addition ($modadd$), and modular inversion ($modinv$) involved in both PADD and PDBL under different coordinate systems. 
% The cost of $modinv$ operation is approximately 100 times that of $modmul$ operation. 

\textbf{PMUL operation.} This operation computes the product of a scalar $s$, a large integer in a finite field, and an elliptic curve point $P$. 
The result, Q, is a new EC point. 
As demonstrated in Algorithm~\ref{alg:danda}, the PMUL operation can be computed by repeated PADD and PDBL operations based on the binary representation of the scalars.
There are two types of point multiplication: fixed point multiplication (FPMUL) and unknown point multiplication (UPMUL). 
For FPMUL, the point P in Algorithm~\ref{alg:danda} is known. The common practice ~\cite{feng2022accelerating, gao2020dpf, mai2019accelerating, pan2017efficient} is to preprocess the known point to eliminate the PDBL operation (line 3 in Algorithm~\ref{alg:danda}). UPMUL, on the other hand, is typically processed faithfully following Algorithm~\ref{alg:danda}.

\begin{wrapfigure}{r}{0.4\textwidth}
\vspace{-0.5cm}
\begin{algorithm}[H]
    \caption{Double-And-Add Algorithm}
    \label{alg:danda}
    \begin{algorithmic} [1]
    \REQUIRE{
        an elliptic curve point $P$, a scalar $s$ with bit-width $l$, $s=(s_{l-1},\dots ,s_1, s_0)_2$; 
    }
    \ENSURE{
        the result point $Q$
    }
    % \begin{algorithmic} [1]
    % \REQUIRE {an elliptic curve point $P$, a scalar $s$ with bit-width $l$, $s=(s_{l-1},\dots ,s_1, s_0)_2$;}
    % \ENSURE{the result point $Q$}
    \FOR{$i \leftarrow 0$ \KwTo $l-1$ \KwSty}
    \IF{$s_i = 1$}
    \STATE$Q = Q+P$
    \ENDIF
    \STATE$P = P+P$
    \ENDFOR
    \end{algorithmic}
\end{algorithm}
\vspace{-0.25cm}
\end{wrapfigure}
\subsubsection{Acceleration for \ecc{}}
As mentioned previously, existing GPU-based solutions~\cite{feng2022accelerating, pan2017efficient, gao2020dpf, mai2019accelerating, huang2020parallel} have mainly focused on reducing the latency of individual PADD and PDBL operations to enhance the throughput of PMUL operations in the Jacobian coordinate system. These solutions typically employ data parallelism, utilizing multi-core processors to process multiple PMUL operations concurrently. 
% However, the PMUL throughput continues to be significantly lower than the actual demand. 
%With the increasing urgency of private data sharing between databases, a higher PMUL throughput is required. 

The solutions above have neglected the potential to enhance the throughput by batching PMUL operations using Montgomery's Trick to reduce the number of $modinv$ operations required in the affine coordinate system, which is one of the contributions of \oursys{} (Section~\ref{sec:batch-ec}).

\subsection{Modular Arithmetic on Finite Field}
\label{sec:background-ff}
\begin{wrapfigure}{r}{0.58\textwidth}
\vspace{-0.5cm}
\begin{algorithm}[H]
    \caption{Montgomery multiplication with SOS strategy}
    \label{alg:ori_montmul}
    \begin{algorithmic} [1]
    \REQUIRE {$a$ and $b$ are stored in A[m:0] and B[m:0] respectively, \\ where $a,b \in F_q$ and $q\_inv \equiv -q^{-1} \text{ mod } D^m$}
    \ENSURE { $c \equiv a*b \text{ mod } q$, store in C[2m:m];}
    \STATE $C[2m:0] = A[m:0] * B[m:0]$ \COMMENT{$\triangleright \ integer \ multiply$}
    \FOR[$\triangleright \ modular \ reduce$] {$x=0 \ to \ m-1$} 
    \STATE $M_i = (C[i] * q\_inv) \ \& \ (D-1);$ \\
    \STATE $C[2m:i] += M_i * q;$
    \ENDFOR
    \RETURN $c > q \ ? \ (c-q) : c;$
    \end{algorithmic}
\end{algorithm}
\vspace{-0.5cm}
\end{wrapfigure}
Finite fields have extensive applications in areas such as cryptography, computer algebra, and numerical analysis, playing a crucial role in \ecc{}. The integer $x$ belongs to a finite field $F_q$, that is, $x \in [0,q)$. Here, $q$ represents a large prime modulus with a bit width $l$ that typically spans 256 to 1024. Generally, the large integer $x$ can be composed as $\sum_{i=0}^{m-1}x_i D^i$, where $D$ symbolizes the base and and $q \le D^m$. $D$ usually set to 64 or 32, allowing the array $X[m:0] =\{x_0 \dots x_{m-1}\}$ to be stored in word-size integers.

A finite field is where the results of $modmul$, $modadd$, and $modinv$ over integers remain within the field. The most time-consuming operations are $modmul$ and $modinv$, which have about 5 times and 500 times the latency of $modadd$ on a mainstream server. These operations are the primary targets of our acceleration efforts.

Montgomery multiplication~\cite{montgomery1985modular} is designed for $modmul$.
% minimize these modular operations. 
By altering the structure of the upper-level loop, various optimization strategies such as SOS and CIOS have been extensively explored~\cite{koc1996analyzing} to reduce read and write operations. As demonstrated in Algorithm~\ref{alg:ori_montmul} (Montgomery multiplication with SOS Strategy), the operation $modmul$ comprises two phases: integer multiplication and modular reduction. The multiplication phase involves multiplying two integers, resulting in a value within the range of $[0, q^2)$. Subsequently, the value is reduced to the interval $[0, q)$ through the modular reduction phase.

As discussed previously, researchers are dedicated to studying the prime moduli of special forms, which provide opportunities to effectively reduce the number of arithmetic operations required for the operation $modmul$. In particular, the use of moduli with characteristics similar to Mersenne primes proves to be beneficial. For instance, the State Cryptography Administration (SCA) of China recommends $q = 2^{256} - 2^{224} - 2^{96} + 2^{64} - 1$, known as SCA-256~\cite{state2016public} which is used in the SM2 curve. 
% Similarly, NIST suggests $q = 2^{256} - 2^{224} + 2^{192} + 2^{96} - 1$, used in P-256 curve.
On the one hand, the exponents in both cases are chosen as multiples of 32, which can expedite implementation on a 32-bit platform. On the other hand, the $q\_inv$ values of the prime in the algorithm~\ref{alg:ori_montmul} are equal to 1, indicating that there is room for optimization in the modular reduction phase.

The $modinv$ operation can be calculated either via Fermat's little theorem (i.e., $x^{-1} \equiv x^{q-2} \text{ mod } q$) or a variant of Extended Euclid Algorithm~\cite{bernstein2019fast, hvass2023high}. The former converts a $modinv$ operation into multiple $modmul$ operations, allowing one to reduce the number of $modmul$ based on known $q$, as demonstrated in the solution~\cite{feng2022accelerating}. The latter employs binary shift, addition, and subtraction operations instead of $modmul$ operations. This results in a performance improvement of about 3 times compared to the former. However, due to the presence of multiple branch instructions within the latter algorithm, it is typically used in CPU-based systems~\cite{OpenSSLNTRU2022} and is not suitable for parallel execution on GPU-based systems.

\begin{wrapfigure}{r}{0.59\textwidth}
    \vspace{-0.25cm}
    \centering
    \setlength{\abovecaptionskip}{0.1cm}
    \includegraphics[width=0.59\textwidth]{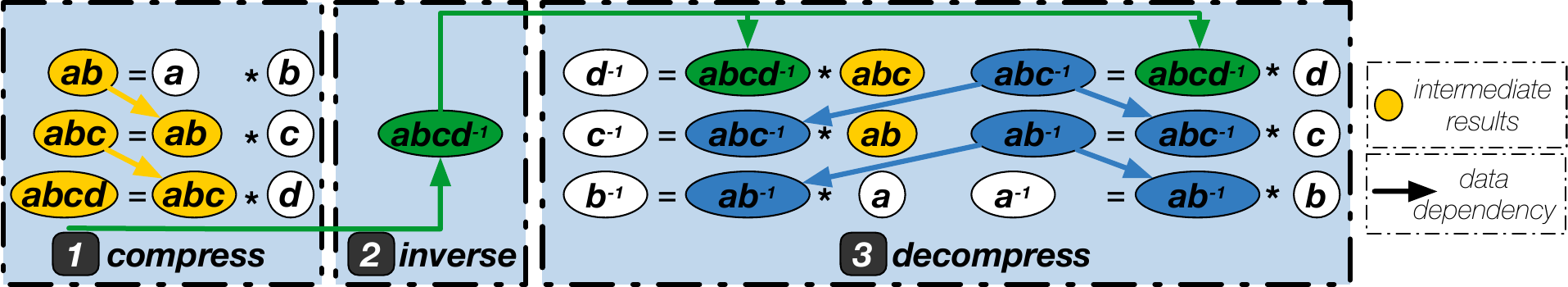}
    \caption{The example of Montgomery's trick.} 
    \label{fig:mont_trick}
    \vspace{-0.25cm}
\end{wrapfigure}
\subsubsection{Montgomery’s trick for batch inversion}
\label{subsec:bg_monttrik}
In this section, we review a well-known method, called \textit{Montgomery's Trick}, which could compress $n$ $modinv$ operations into one $modinv$ operation by introducing $3n$ $modmul$ operations. There are three key steps in Montgomery's trick. For example, to calculate $a^{-1}, b^{-1}, c^{-1}, d^{-1}$ from the inputs $a, b, c, d$, as shown in Fig.~\ref{fig:mont_trick}, the first step, called \textit{compress step}, merges these inputs through cumulative multiplications. Then we calculate the inverse of $abcd$ to get $abcd^{-1}$, called the \textit{inverse step}. Finally, we could get $a^{-1}, b^{-1}, c^{-1}, d^{-1}$ by reversing the compression process, called the \textit{decompress step}. 

% \begin{equation}
% \label{form:compress}
% \begin{aligned}
% ab &= a * b \\
% abc &= ab * c \\
% abcd &= abc * d
% \end{aligned}
% \end{equation}

% \begin{equation}
% \label{form:uncompress}
% \begin{aligned}
% d^{-1} &= abcd^{-1} * abc &\quad abc^{-1} &= abcd^{-1} * d \\
% c^{-1} &= abc^{-1} * ab &\quad ab^{-1} &= abc^{-1} * c \\
% b^{-1} &= ab^{-1} * a &\quad a^{-1} &= ab^{-1} * b
% \end{aligned}
% \end{equation}

\subsubsection{Acceleration for Modular Arithmetic}
Extensive research has been conducted on accelerating $modmul$ operation on GPUs, falling into two categories: one that uses a mixture of multi-precision floating-point instructions and integer instructions, and the other using only integer instructions. Since the introduction of the independent floating point unit in Nvidia's GPU Pascal architecture, many studies~\cite{gao2020dpf, gao2021dpf, emmart2018faster, dong2018sdpf} have replaced integer multiplication with multi-precision floating-point multiplication. Among them, the fastest implementation is based on $D=52$, called DPF. Efforts persist to implement $modmul$ using only integer arithmetic instructions, with notable studies~\cite{zheng2014exploiting, cgbn, dong2018towards, sppark2024} like CGBN and sppark, which were fastest in the ZPrize competition~\cite{zprize}, though sppark does not support 256-bit prime moduli due to carry propagation issues. % There is a long-term effort to implement $modmul$ using only integer arithmetic instructions. There are many related studies~\cite{zheng2014exploiting, cgbn, dong2018towards, sppark2024}, which are implemented at the assembly level using Parallel Thread Execution (PTX)~\cite{ptx2024}. Among them, CGBN~\cite{cgbn} and sppark~\cite{sppark2024} are confirmed to be the fastest in the ZPrize competition~\cite{zprize}. However, sppark does not support 256-bit prime moduli due to its carry propagation issue. 
From a parallelism perspective, the implementations of $modmul$ operation can be single-threaded or multi-threaded. The latter utilizes inter-thread communication warp-level primitives~\cite{cuda2024} to complete a $modmul$ operation collaboratively. For instance, CGBN and DPF support setting 1, 2, 4, 16, and 32 threads.

%% file: alg_layer.tex
\section{Throughput-oriented Elliptic Curve Operations}
\label{sec:batch-ec}
In this section, we describe the design of throughput-oriented \ec{} operations based on the affine coordinate with Montgomery's Trick.

% public-key cryptography: ECDSA， PSI的计算过程，
% 抽取共同的计算块（算法展示？），
% 已知s*P的运算，高吞吐为导向的性能需求，加速目标是提高ops/s？要解决的问题是什么？
% \para{Profiling or observations.}
% % 这里是否可以给出模逆运算和模乘运算的性能差距，以及实际应用场景中的计算请求并发数量来促成我们的motivation？
% % 1. 性能对比 模逆：模乘=1:100？用模乘来量化批量请求的计算总量
% % 2. 批量模逆运算可以montgomery trick来转换为少量的模逆和一系列存在依赖的模乘运算
% % \para{Insight.}
% % insight: 
% % 1. 批量point mul运算，使得affine coordinate with Montgomery Trick成为可能，降低总的计算量； inverse operation is affordable 
% % 2. 传统double and add算法，compute-bound，“batch add算法降低计算量的同时引入更多数据读写，可通过计算与数据传输的overlap来缓解compute bound的性能损失？”

% \para{Challenges.}
% % challenges：
% %% 1. batch point addition 引起的数据读写问题/cache不足？
% %% 2. scalar差异导致warp branch divergency问题？我们是否针对这个问题减轻了一些computation overhead呢？
% % 延后读取scalar值并进行判断是否更有意义？

% 创新：
%% 1. batch point addition/dbl
%% opt1. batch point addition to reduce computation overhead: algorithm or calculation redesign, 需要解释当batch size没有达到阈值时，通过data padding（在计算数据中补充one值的方式，凑齐batch size），引起的“额外计算”开销是否影响性能？
%% cuda opt1: optimze memory access, only read nessaray data, for example only read the x of each point， the needed scalar data， and so on
%% cuda opt2: recompute, instand of storing and reading the old data
%% cuda opt3: data prefetch
%% cuda opt3: avoid warp divergence，thread branch problem
%% 2. fixed point mul
%% 2.1 only use batch point addition algorithm? 在此基础上可根据计算特征进一步优化性能，
%% 2.2 沿用已有的数据预处理（P，2p, 3p, xxx）优化，加速性能：
%%
%% 3. unfixed point mul
%% 3.1. batch point dbl 和 batch point addition kernel
%% 3.2. kernel fusion to reduce memory access, improve 10% additional performance 
%% 

\subsection{Opportunities and Challenges of Batching EC Operations}
\label{subsec:calculation-ana}
%Recent reports indicated that there are billions of digital signature generation or verification requests generated per second~\cite{rescorla2018transport, pinkas2019spot}.
% Formally, we consider a request set $R = \{r_1, r_2, ..., r_N\}$, where each $r_i$ contains a few PMUL operations ${Q_i} = {s_i}*{P_i}$. 
% Each PMUL operation contains hundreds of PADD operations and PDBL operations.
% Specifically, a signature verification has one FPMUL and one UPMUL operation. Each signature generation contains a $modinv$ and an FPMUL operation.
% These operations are the most time-consuming primitives in the ECDSA.
% The existing works~\cite{feng2022accelerating, gao2020dpf} focus on reducing the latency of PMUL operation and $modinv$ by leveraging its mathematical properties.
% However, they ignore the opportunity to improve the throughput of these primitives further.

%\cb{The PMUL operation contains hundreds of PADD operations and PDBL operations.
%The existing works~\cite{feng2022accelerating, gao2020dpf} focus on reducing the latency of PADD operation and PDBL operation to improve the performance of PMUL operation.
%However, they ignore the opportunity to improve the throughput of these primitives further.}

Our key hypothesis is that we can batch and execute \ec{} operations in parallel to enhance throughput.   
% Our key insight is that collaboratively processing batch elliptic curve operations could be faster. 
% When processing batch PMUL operations in parallel, a large number of PADD operations and PDBL operations are computed simultaneously.
The overhead of processing a number of primitive operations (PADD and PDBL) could be reduced by employing a more efficient method based on the affine coordinate system.
For $N$ PADD operations, \oursys first calculates $\lambda$ (Eqn.~\ref{form:point-add-1}) of each PADD by one $modmul$ operation and one $modinv$ operation.
Specifically, \oursys batches all the $modinv$ operations in multiple PADD operations together and processes them with Montgomery's trick.
The $N$ $modinv$ operations are converted into one $modinv$ operation and $3N$ $modmul$ operations.
After getting the $\lambda$ of each PADD operation, \oursys only needs another 2 $modmul$ and 6 $modadd$ operation by leveraging  Eqn.~\ref{form:point-add-2} to compute the output.
In summary, there are about one $modinv$, $6N$ $modmul$, and $6N$ $modadd$ operations for batch $N$ PADD operations with affine coordinates, while there are $11N$ $modmul$ and $6N$ $modadd$ for $N$ PADD operation with Jacobian coordinates, as shown in Table~\ref{tab:coord-select}.
% The cost of $modinv$ operation is approximately $100$ times that of $modmul$ operation.
Although a few expensive $modinv$ operations are involved, by amortizing these costs through large batch sizes, the total overhead of batch PADD operations could be reduced.
Theoretically, when the batch size $N$ is greater than 20, the total overhead of processing a batch of $N$ PADD operations in the affine coordinate is less than that in the Jacobian coordinate system.
The actual size of $N$ is much larger than $20$ to fully utilize the large-scale parallelism provided by GPUs.
\oursys accelerate the processing of a batch of PDBL operations in a similar fashion. %and hence can eventually accelerate a batch of PMUL.
%At last, the batch PMUL operations could be also accelerated based on the batch PADD / PDBL operations.
%对比基于雅可比坐标的padd并发计算复杂度和基于affine坐标的padd并发计算复杂度（实际计算量）
% Considering $N$ point addition operations, there are $11N$ modular multiplications to calculate these PADDs based on the jacobian coordinate system.
% We use $M$ to denote the number of streaming multiprocessor(SM) in GPU. 
% Considering the expensive overhead of modular inversion, we need to batch process enough point operations to reduce the overall computational cost. 
% And we conduct a computational analysis to find the lower threshold for the scale of batch point operations. 

% 给出实现batch PMUL面临哪些挑战
% 1. batch inv的串行依赖导致的并行性问题
% 2. 由于需要足够的batch大小，才能降低总的计算开销，从而导致的点数据重复读取问题，带来显著的数据访问开销；或者换个角度，从存储开销问题入手，引出数据访问开销，感觉也可以
Although this idea sounds promising, it is non-trivial to capitalize on its benefits due to the following two challenges.

\begin{figure*}[h]
    \centering
    \vspace{-0.25cm}
    \setlength{\abovecaptionskip}{0.01cm}
    \subfigure[The throughput bottleneck on each GPU Steaming Processor with data parallelism]{
        \includegraphics[width=0.9\linewidth]{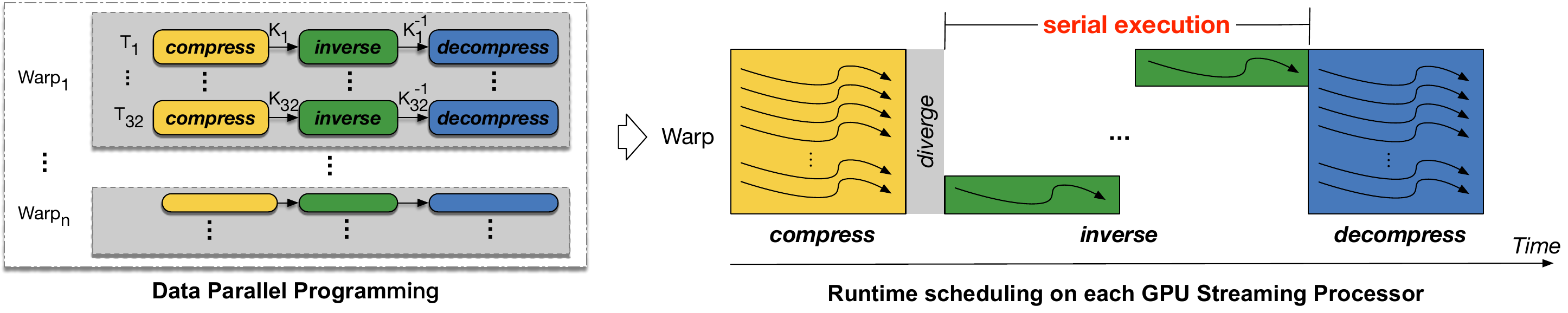}
        \label{fig:cgpw1}
    }
    \subfigure[\sys{'s} runtime execution on each GPU Steaming Processor with GAS mechanism]{
        \includegraphics[width=0.9\linewidth]{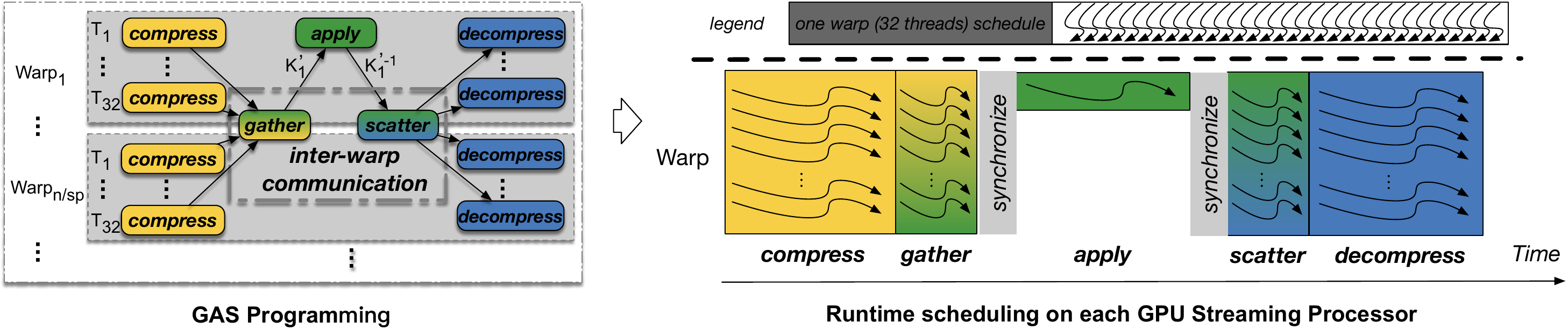}
        \label{fig:cgpw}
    }
    \caption{Different mechanism of batch modular inversion on GPU and corresponding runtime execution.} 
    \vspace{-0.25cm}
\end{figure*}

\textbf{Challenge 1.} The calculation of the batch $modinv$ with Montgomery's trick is inherently sequential since $modinv$ needs the results of the multiple accumulated $modmul$ operations. 
% The data parallel model can be used to accelerate batch $modinv$, as illustrated in the left part of Figure~\ref{fig:cgpw1}. \cb{The most fundamental scheduling unit in a GPU is a warp, which comprises 32 threads. These threads run concurrently on a single GPU streaming processor, which houses multiple CUDA cores. For instance, NVIDIA A100 contains sp=432 streaming processors. }
Although this shortens the sequential dependency in the compress/decompress step, we have to introduce more time-consuming $modinv$ operations. Moreover, the large number of cores on a GPU requires a high degree of data parallelism to fully utilize its computing power, which results in introducing many more $modinv$ calculations.
Even worse, these $modinv$ operations are computed serially rather than in parallel on the GPU Streaming Processor (SP), as demonstrated in the right part of Fig.~\ref{fig:cgpw1}.
The reason is that the fastest algorithm~\cite{bernstein2019fast, hvass2023high} for $modinv$ employs numerous branching conditions to achieve efficient computation, which causes a large number of GPU warp divergences in parallel computing and then results in worse performance.
%Our investigation shows that the proportion of the \textit{inverse step} (Fig. \ref{fig:mont_trick}) in the native data parallel model for batch $modinv$ is up to $80\%$ when the input's scale is $2^{20}$.
%The significant latency of the $modinv$ operation further exacerbates the overhead of this part.
To address this, we carefully devise a parallel workflow for Montgomery’s Trick algorithm in Section~\ref{subsec:pminv} to minimize the number of the \textit{inverse step} of $modinv$ computation and improve the efficiency of parallel computing.

\textbf{Challenge 2.} When processing a batch of \ec{} operations using Montgomery’s Trick, there are much higher memory access overheads compared to existing methods, such as~\cite{feng2022accelerating}.
Specifically, in the \textit{compress step} (Fig.~\ref{fig:mont_trick}), we need to load two EC points for each PADD operation to calculate the numerator and denominator of $\lambda$, and then multiply the denominator of $\lambda$ of different PADD operations together to compress the $modinv$ operations needed and store them for subsequent calculation of the \textit{decompress step} (Fig.~\ref{fig:mont_trick}).
Suppose that we process a batch of millions (e.g., $2^{20}$) of EC operations to maximize the parallelism. The point data and intermediate arrays would consume more than \SI{128}{MB} \SI{96}{MB}, respectively.
However, this far exceeds the capacity of L1 cache or shared memory of modern GPUs. For example, the combined L1 cache and shared memory in NVIDIA A100 GPU has only \SI{20}{MB}.
%We have to store the intermediate results in the large GPU's global memory in the \textit{compress step} and reload them in the \textit{decompress step} to compute the value of $\lambda$.
%Then the point data also stored in GPU global memory is loaded again to finish the complete calculation of PADD operation.
Offloading these data to global memory would incur huge overhead for reloading them to the cache for computation in both the \textit{compress step} and the \textit{decompress step} Fig.~\ref{fig:mont_trick}. 
% In addition, there is little opportunity to hide this overhead by overlapping the data loading and computation.  
%significantly decreasing the performance of batch EC operations.
To tackle this issue, we propose two significant optimizations: first, a memory management mechanism for batch PADD operations that actively utilizes a multi-level cache to cache data (see Section~\ref{subsec:batch-padd}); second, a data-locality-aware kernel fusion method for batch PMUL operations, which aims to reduce memory access frequency and enhance data locality (Section~\ref{subsec:batch-pmul}).

% 针对挑战1:
% 1. 并行方案：压缩模逆运算次数，牺牲模逆运算的部分并发度（block-level， not warp-level），消除warp divergency issue
%% montgomery's trick 
\subsection{Batch Modular Inversion Optimization}
\label{subsec:pminv}
\label{subsec:batch-inv}

% \para{Summary: coarse-grained parallel for batch modular inversion. }
% We propose the partitioned parallel workflow for batch modular inversion, which not only ensures the fast reduction of large scale multiplication operations, but also maximizes the parallelism of batch modular inversion while avoiding warp divergence issue.
Here, we introduce the design of batch $modinv$ operation in \oursys{}, which serves as a crucial component of the throughput-centric \ec{} operations. 
In a GPU, the fundamental scheduling unit is a warp that consists of 32 threads. These threads run concurrently on a single GPU SP, which houses multiple CUDA cores. For example, NVIDIA A100 contains 432 SPs. 
% To ensure sufficient overlap between computation and memory access overhead, multiple warps are typically run on a single GPU streaming processor. As presented in Section~\ref{sec:background-ff}, due to the presence of multiple branch instructions in the fastest $modinv$ operation based on the Extended Euclid Algorithm, there is divergence among the threads within a warp during the inversion step, as shown in Figure~\ref{fig:cgpw1}. This leads to serial execution at runtime, resulting in a significant performance drop. Therefore, reducing $modinv$ operations of the \textit{inverse step} is the focus of acceleration.

% We propose a partitioned parallel method with Montgomery’s trick algorithm for batch $modinv$, which ensures the fast reduction of large-scale modular multiplications and maximizes the parallelism of modular inversion while avoiding warp divergence.
% \oursys uses a divide-and-conquer strategy to split the large scale of inputs into multiple parts, and then assigns each part to a GPU block for parallel processing.
% The figure~\ref{fig:cgpw} illustrates how to parallel large-scale modular inversions in GPU.
% \oursys exploits the GPU inter-block Gather-Apply-Scatter (GAS) mechanism to map the inputs into GPU threads further for parallel processing and minimize the needed modular inversions through inter-block reduction.
To address \textbf{Challenge 1}, we borrow the wisdom of parallel graph processing systems and adopt the \textbf{Gather-Apply-Scatter (GAS)} mechanism~\cite{joseph2012powergraph} to reduce the overhead of the $modinv$ operation. Similar to data parallelism, we evenly distribute the $N$ inputs that require inversion across $32n$ threads in $n$ warps for parallel processing as shown in the left part of Fig.~\ref{fig:cgpw}. Each thread then executes the \textit{compress step} to obtain the final accumulated products, $K_i$ ($1\leq i \leq 32n$). The difference lies in the fact that each thread does not directly compute the \textit{inverse step} to get $K_i^{-1}$. Instead, it waits for the \textbf{GAS} synchronization to complete, obtaining $K_i^{-1}$, and then performs the \textit{decompress step}.

In the \textbf{gather} phase, the accumulated products $K_i$ from all threads in the $n/sp$ warps running on each GPU SP are collected. After applying the \textit{compress step} again, we obtain the accumulated products $K_j^{'}$ (where j ranges from $1$ to $sp$, and $sp$ is the number of GPU streaming processors). For instance, the accumulated products $K_i$ (where i ranges from $1$ to $32n/sp$) from $Warp_1$ to $Warp_{n/sp}$ are compressed using shared memory for inter-thread communication, yielding the result $K_1^{'}=\Pi_{i=1}^{32n/sp}K_i$, as shown in the left part of Fig.~\ref{fig:cgpw}.

During the \textbf{apply} phase, each GPU SP performs a $modinv$ operation in the \textit{inverse step} on $K'_j$ using one thread, resulting in $K_j^{'-1}$, as shown in the right part of Fig.~\ref{fig:cgpw}. This effectively avoids the divergence issue caused by multiple threads performing the $modinv$ operations simultaneously. Although some CUDA cores in the GPU SPs are idle during this time, this period is very short relative to the entire computation process.
In the \textbf{scatter} phase, the $K_j^{'-1}$ obtained by each GPU SP is decompressed to produce the corresponding $K_i^{-1}$.

Compared to a naive data parallel approach, batch inversion based on the \textbf{GAS} mechanism reduces the number of required $modinv$ operations from $32n$ to $sp$, where $n$ is typically several times the value of $sp$ to use computing to overlap the memory access overhead. This effectively reduces the overhead of batch inverse.

\subsection{Batch Point Addition Optimization}
\label{subsec:batch-padd}
\begin{figure}[h]
    \vspace{-0.25cm}
    \centering
    \setlength{\abovecaptionskip}{0.1cm}
    \includegraphics[width=0.9\textwidth]{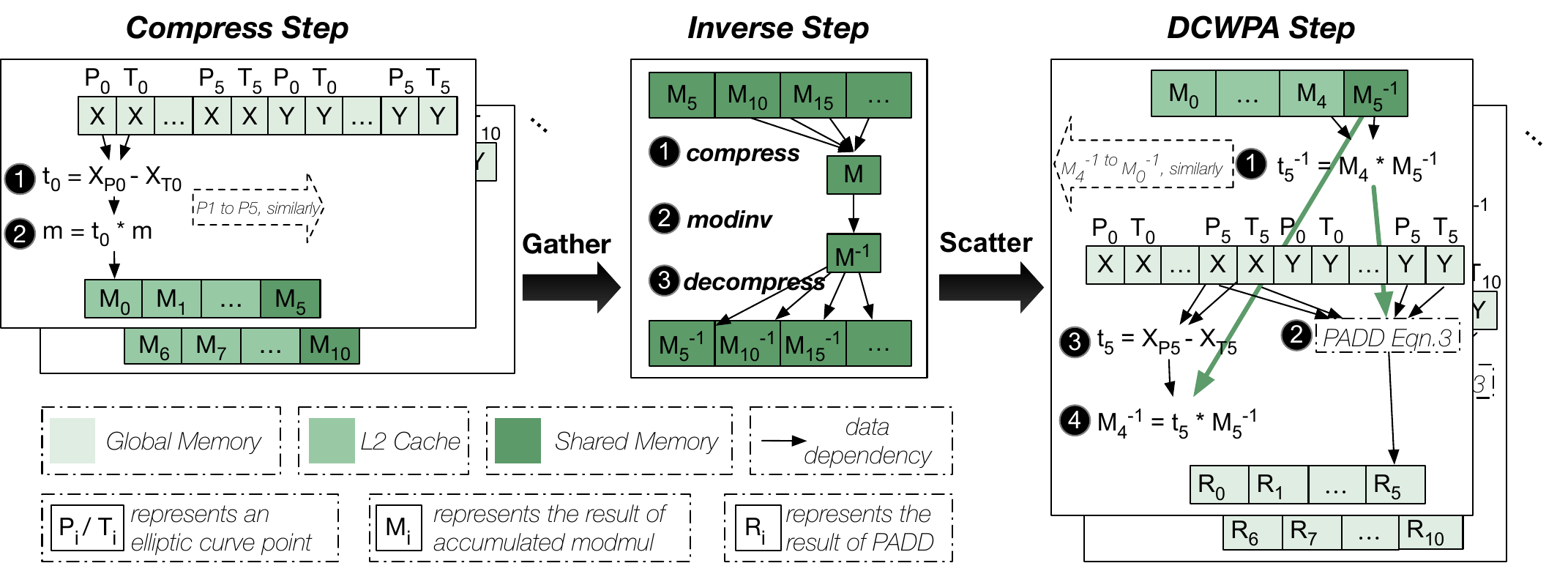}
    \caption{Multi-level cache management for batch PADD.} 
    \label{fig:bpadd-mco}
    \vspace{-0.25cm}
\end{figure}

For batch PADD operation, \oursys first calculates the $\lambda$ for all PADD operations using batch $modinv$ and then completes the computation of Eqn.~\ref{form:point-add-2} of each PADD, which are combined with the \textit{decompress step} of batch $\lambda$ to reuse the point data.
We called this combined \textit{decompress step} as \textit{DCWPA step}.
As we discussed in \textbf{Challenge 2}, there are high memory access overheads due to
frequent global memory access.
To alleviate this, \oursys employs multi-level cache management to minimize data access overhead, as shown in Fig.~\ref{fig:bpadd-mco}.
\oursys reduces the overhead of memory access with two major techniques. %: reducing the number of memory accesses and improving memory access efficiency.
1) We minimize caching the intermediate data produced in the \textit{compress step} and recompute them when needed in the \textit{DCWPA step}, decreasing the required cache space. %in the \textit{DCWPA step}.
2) \oursys classifies the data according to the computing characteristics and allocates them at the optimized cache level, improving the data access efficiency.

To minimize the number of memory accesses when processing PADD operations, \oursys simplifies the computation of the \textit{compress step}.
The core idea is that \oursys only accesses the data needed to get the modular inverse of the denominator of $\lambda$.
As shown in the \textit{compress step} in Fig.~\ref{fig:bpadd-mco}, we only load each EC point's $X$ coordinate value to get the denominator $t_i$ of $\lambda$ and store the intermediate result $m_i$ of the accumulated products to the array $M$ in the \textit{compress step}.
The nominator of $\lambda$ is only used to get the result of the PADD, so we calculate it in the \textit{DCWPA step}.
It should be noted that \oursys does not store the denominator and the nominator of $\lambda$ to the intermediate array $M$ for each PADD but recomputes them in the \textit{DCWPA step}.
The reason is that $modadd$ operations are cheaper than global memory access of a large integer.
Furthermore, the \textit{DCWPA step} has to load the complete EC point data anyway; therefore, the recomputation does not incur additional data loading overhead.
As shown in the \textit{DCWPA step} in Fig.~\ref{fig:bpadd-mco}, the operation \ding{204} recalculates the denominator $t$ of $\lambda$ based on the loaded point data.
Therefore, \oursys reduces the intermediate data stored by introducing a little computation overhead.
This scheme is also applicable to PDBL operations.

In addition, \oursys assigns the different data to different caches to align with the computation.
We observe that the data generated later in the \textit{compress step} are used earlier in the \textit{inverse step} and \textit{DACWPA step}.
For example, the last result $M_5$ of the accumulated products for each thread in the \textit{compress step} is immediately used for the \textit{inverse step}, and its inverse value ${M_5}^{-1}$ is first used in the \textit{DCWPA step}.
To support the \textit{gather phase} in the \textit{inverse step}, \oursys utilizes GPU shared memory to cache the last result of the accumulated products for all threads in a GPU block in the \textit{compress step}, as shown in the \textit{inverse step} in Fig.~\ref{fig:bpadd-mco}.
The \textit{gather phase} requires frequent data access to merge data from different threads in one block.
GPU's shared memory could meet this demand by enabling data communication between threads and supporting simultaneous access by multiple threads.
Moreover, the shared memory of Nvidia GPUs allows for very low memory access latency, thus reducing memory access overhead in the \textit{inverse step}.

\oursys also leverages the L2 persistent cache to store all intermediate data of accumulated products in the \textit{compress step}.
As demonstrated in Fig.~\ref{fig:bpadd-mco}, the array $M$, stored the intermediate result of each accumulated product in the \textit{compress step}, is actively cached in the L2 persistent cache of Nvidia A100 GPU. 
As we discussed in Section~\ref{subsec:calculation-ana}, the intermediate arrays will occupy tens of megabytes of memory space.
Although \oursys has reduced the stored intermediate data, the needed memory space still far exceeds the limited shared memory space.
For example, the stored data in the \textit{compress step} is \SI{32}{MB} when the scale of batch PADD operations is $2^{20}$.
Besides, in the last \textit{DCWPA step}, we need to first access the array $M$ to generate the inverse value of the denominator of $\lambda$.
To meet these requirements, \oursys uses the L2 persistent cache with larger capacity and lower access latency to optimize these data accesses.
The Nvidia Ampere architecture supports the L2 cache residency control feature~\cite{nvidiaA100}, allowing users to customize the large \SI{40}{MB} L2 cache more efficiently.
According to the CUDA programming guide~\cite{cuda2024}, users can configure to use up to $75\%$ L2 cache.
\oursys sets aside a portion of the L2 cache as the persistent cache and configures the array $M$ on it, thus achieving lower latency accesses. 
When the allocated space exceeds the configured L2 persistent cache, \oursys uses the first-in-first-out (FIFO) cache replacement policy provided by CUDA to align with the calculation of the \textit{DCWPA step}. 
We know that the last result of multiplication in the \textit{compress step} is first used in the \textit{DCWPA step} to compute the corresponding inverse value.
The FIFO cache replacement strategy is perfectly suited to such a computational feature, improving the hit rate of the L2 cache.
Finally, the original point data array (\eg, $P_i$ and $T_i$ in Fig.~\ref{fig:bpadd-mco}), which occupies several hundred megabytes of space, is stored in the GPU's global memory.
% \oursys carefully reorders the calculation process in the \textit{DCWPA step} to maximize the overlap between global memory accesses and the large integer computations, minimizing the memory access overhead.

\begin{wrapfigure}{r}{0.58\textwidth}
    \vspace{-0.5cm}
    \centering
    \setlength{\abovecaptionskip}{0cm}
    \label{fig:dl}
    \subfigure[The row-majored data layout]{
        \includegraphics[width=0.2\textwidth]{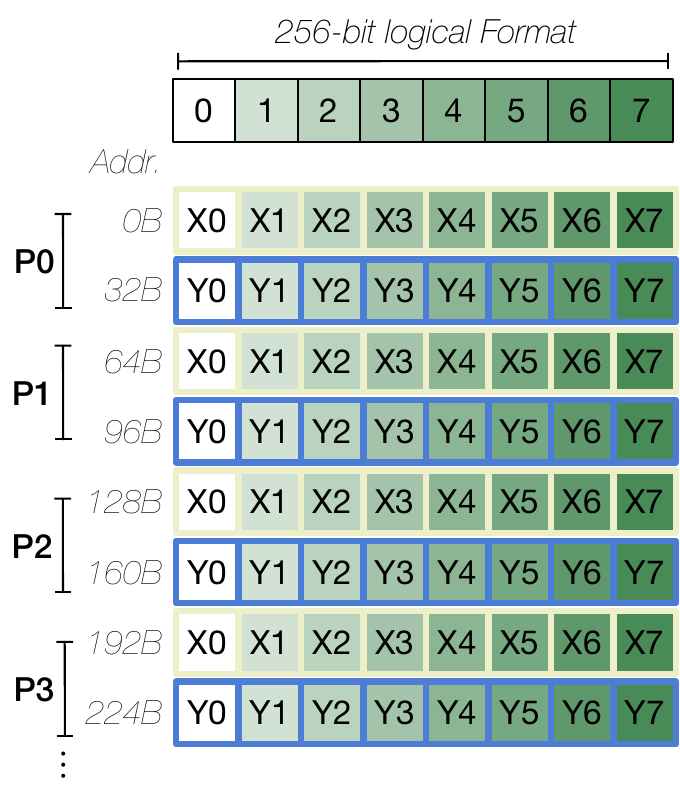}
        \label{fig:rmdl}
    }
    \subfigure[The column-majored data layout]{
        \includegraphics[width=0.35\textwidth]{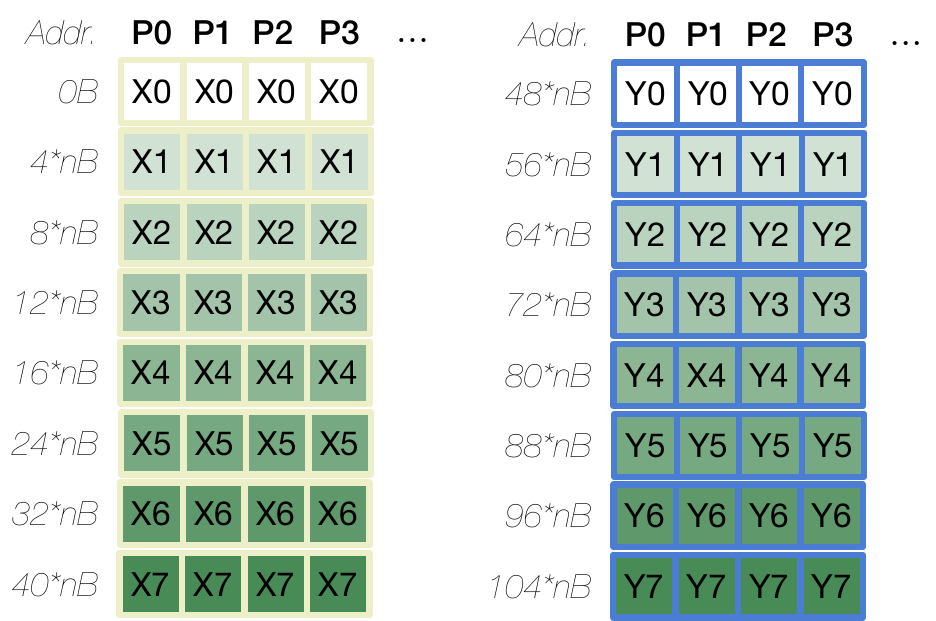}
        \label{fig:cmdl}
    }
    % \Description{xxx}
    \caption{The data layout of n EC point}
    \vspace{-0.25cm}
\end{wrapfigure}
\para{Cache-efficient data layout.}
\oursys adopts a cache-efficient column-majored data layout for large integers to achieve efficient concurrent data access, improving the efficiency of global memory access.
% The input for the PMUL operation includes a scalar value $s$ and an elliptic curve point $P$, which are both large prime integers.
Usually, a large integer is represented by an integer array in a CPU, as mentioned in section~\ref{sec:background-ff}. %Multiple large integers are stored contiguously in the physical memory space with a row-majored data layout.
As illustrated in Fig.~\ref{fig:rmdl}, the two coordinate values (X, Y) of the EC point $Pi$ are stored in the two integer arrays $X$ and $Y$, respectively, each element occupying 32 bits. In the row-majored data layout, $X$ and $Y$ of an EC point occupy a contiguous physical memory space.
However, this row-majored data layout is unsuitable for GPU concurrent access.
When the threads in a GPU warp load the point data, each thread $Ti$ serially accesses the element of the array $X$ first.
As shown in Fig.~\ref{fig:rmdl}, the stride between threads is up to 64 bytes, which means there is no aggregate access in a GPU warp for each large integer access and causes long data access latency.
To address this issue, \oursys separates the X and Y coordinates of the point data and places the data of the same type together in a \textit{column-majored data layout}, such as the X coordinates of all the points in Fig.~\ref{fig:cmdl}.
For example, we store the first element $X0$s of all X-axis large integers contiguously, then all the second elements, until the last one $X7$.
Now, the data accessed by threads within a warp is contiguously located, supporting efficient aggregation access.
In addition, we eliminate the additional overhead of data transpose by overlaying the transpose of the next batch of data with the calculation of the current batch.

\subsection{Batch Point Multiplication Optimization}
\label{subsec:batch-pmul}
According to the double-and-add algorithm~\ref{alg:danda}, for each bit of the scalar, a PADD operation is needed when the current bit of the scalar is one and there is a PDBL operation after the PADD operation when the loop is from the least significant bit of the scalar to the most significant bit.
The input of the PDBL is irrelevant to the output of the PADD, and its output is used to calculate the next loop.
Based on this, \oursys proposes data-locality-aware kernel fusion optimization for batch PMUL operations.
The goal of our design is to reduce the memory access frequency and enhance data locality.

\begin{figure}[t]
    \vspace{-0.25cm}
    \centering
    \setlength{\abovecaptionskip}{0.1cm}
    \includegraphics[width=0.9\linewidth]{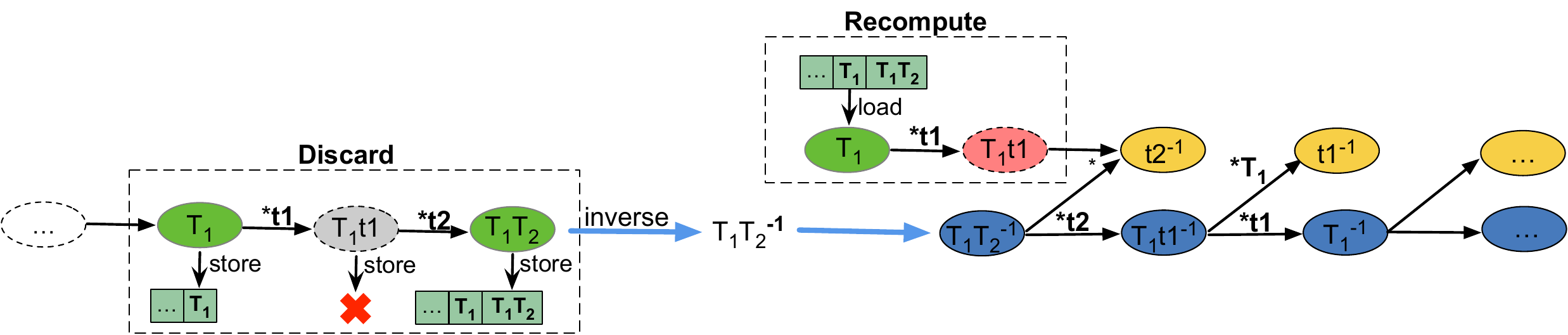}
    \caption{An example of find-then-recompute method.} 
    \label{fig:find-and-recom}
    \vspace{-0.25cm}
\end{figure}

Initially, the three steps of batch PADD operation are combined with the corresponding steps of batch PDBL operation.
This native fusion could improve the data locality by reusing the loaded point data but requires more memory space to store the intermediate data.
For example, two intermediate arrays (\ie, the array $M$) are necessary for each thread in the combined \textit{compress step}:
one for the PADD operation and the other for the PDBL operation.
The next \textit{inverse step} could work on both of them, simply doubling the input size and further merging the inverse calculations into one. 
\begin{wrapfigure}{r}{0.5\textwidth}
% \vspace{-0.5cm}
\begin{algorithm}[H]
    \begin{algorithmic} [1]
    \caption{Batch UPMUL operation}
    \label{alg:batch-pmul}
    \REQUIRE{
        the (scalar, point) set \{$(s_{0}, P_{0}), (s_{1}, P_{1}), ..., (s_{(n-1)}, P_{(n-1)})$\}
    }
    \ENSURE{
        \{$Q_0, Q_1, Q_2, ..., Q_{n-1}$\}
    }
    \STATE $len\_s \leftarrow$ the bits of scalar; \\
    \STATE $M[n]$; \\
    \FOR{$i \leftarrow 0$ \KwTo $len\_s - 1$ \KwSty} 
        % $mul\_acc = 1$; \\
        % \For{$j \leftarrow 0$ \KwTo $n-1$ \KwSty}{
        %     $M[j] = mul\_acc;$ \\
        %     % $mul\_acc\_chain[i] = mul\_acc;$ \\
        %     % \tcc{calculate the denominator of $\lambda_{PDBL}$} 
        %     % \tcc{load the y-axis of point $P_i$ from global memory} 
        %     $t1 = P_{j}.Y + P_{j}.Y;$ \\ 
        %     $mul\_acc = mul\_acc * t1;$\\
        %     % \tcc{calculate the denominator of $\lambda_{PADD}$} 
        %     $t2 = P_{j}.X - Q_{j}.X;$ \\ 
        %     $mul\_acc = mul\_acc * t2;$\\
        % }
        % ${mul\_acc}^{-1} = inverse\_step\_func(mul\_acc)$; \\
        \STATE $T_1 = 1$; \\
        \FOR{$j \leftarrow 0$ \KwTo $n-1$ \KwSty}
            \STATE $M[j] = T_{1};$ \\ 
            \STATE $t1 = P_{j}.Y + P_{j}.Y;\ T_{1}t1 = T_1 * t1;$\\
            % \tcc{calculate the denominator of $\lambda_{PADD}$} 
            \STATE $t2 = P_{j}.X - Q_{j}.X;\ T_{1}T_{2} = T_{1}t1 * t2;$ \\ 
            \STATE $T_{1} = T_{1}T_{2};$
        \ENDFOR
        \STATE ${T_{1}}^{-1} = inverse\_step\_func(T_{1})$; \\
        \FOR{$j \leftarrow n-1$ \KwTo $0$ \KwSty}
            \STATE$T{_1}t1 = (P_{j}.Y + P_{j}.Y) * M[j];$ \\
            \STATE${t2}^{-1} = {T_{1}}^{-1} * T{_1}t1;$ \\
            \STATE $\lambda_{PADD} = \lambda_{PADD}\_func(P_j, Q_j, {t2}^{-1});$ \\
            \STATE $R_j = padd\_func(\lambda_{PADD},\ P_j,\ Q_j);$ \\
            \STATE ${T_{1}}^{-1} = {T_{1}}^{-1} * (P_{j}.X - Q_{j}.X);$ \\
            \STATE ${t1}^{-1} = {T_{1}}^{-1} * M[j];$ \\
            \STATE $\lambda_{PDBL} = \lambda_{PDBL}\_func(P_j, {t1}^{-1});$ \\
            \STATE ${T_{1}}^{-1} = {T_{1}}^{-1} * (P_{j}.Y + P_{j}.Y);$ \\
            \STATE $P_j = pdbl\_func(\lambda_{PDBL},\ P_j);$ \\           
            \IF{$s[i] == 1$}
            \STATE $Q_j = R_j$\\
            \ENDIF
        \ENDFOR
    \ENDFOR  
    \end{algorithmic}
\end{algorithm}
\vspace{-0.25cm}
\end{wrapfigure}
However, the limited shared memory will limit the input scale, thus decreasing the throughput of batch PMUL operations.

To achieve the design goal, \oursys uses a \textit{ find-then-recompute} method to fusion the two kernels, which could reduce memory space and improve data locality.
There is an example of applying the find-then-recompute method to Montgomery's trick, as illustrated in Fig.~\ref{fig:find-and-recom}.
This method discards part of the results of the accumulated multiplications to reduce the space occupied by array $M$, then recomputes the discarded data with the intermediate result stored and the original inputs.
For example, when multiplying $t1$ and $t2$ with the previous accumulated product result $T_1$, we discard the result ${T_1}t1$ of multiplying $t1$ and $T_1$ and only store the final result ${T_1}{T_2}$ of the product of $t1$, $t2$, and $T_1$.
Then, to get the correct inverse values of $t1$ and $t2$ in the \textit{decompress step}, these values ${T_1}$, ${T_1}t1$, and ${T_1}{T_2}^{-1}$ are needed according to Montgomery's trick.
We could recompute the intermediate result ${T_1}t1$ based on the stored value $T_1$ and the input $t1$ to complete the remaining computation.

Specifically, \oursys uses the same variable $T_1$ to multiply with the denominator of $\lambda_{PADD}$ and the denominator of $\lambda_{PDBL}$ together and store the last result in the intermediate array $M$ as shown in the lines $6-9$ in Algorithm~\ref{alg:batch-pmul}.
In addition, \oursys adjusts the order of the calculation to achieve overlap between global memory access and large integer computation to reduce latency. Firstly, we only load the y-axis value of the point $P_{j}$ to calculate the denominator of $\lambda_{PDBL}$.
By leveraging the cache-efficient data layout, \oursys could aggregate data access to any part of a large integer at a smaller granularity (32-bit). 
% More details are in Section~\ref{subsec:lamm}.
When calculating the denominator of $\lambda_{PDBL}$ and then multiplying by $T_1$, \oursys loads the values on the x axis of the point $P_{j}$ and $Q_{j}$ to overlap the long latency of global memory access with the large integer multiplication. 
Then, we calculate the denominator of $\lambda_{PADD}$ and multiply by $T_1$, Finally, the value of $T_1$ for the point pair ($P_{j}, Q_{j}$) is stored in the array $M$.
Next, the calculation of the \textit{inverse step} is the same as the one in batch PADD operations. \oursys also reduces the latency of the \textit{inverse step} compared to the native kernel fusion method.

In the \textit{DCWPA step}, we first complete the calculation of PADD operation and then the calculation of PDBL operation.
\oursys gets the inverse value of the denominator of $\lambda_{PADD}$ by recomputing the discarded values, the denominator $t1$ of $\lambda_{PDBL}$ and the value $T{_1}t1$, as shown in the lines $13-15$ in Algorithm~\ref{alg:batch-pmul}.
Then, it completes the computation of Eqn.~\ref{form:point-add-2} to obtain the result $R_j$ of PADD operations and updates the value $Q_j$ using $R_j$ according to the j-th bit value of the scalar.
At last, just like handling PADD operation, \oursys calculates the value of $\lambda_{PDBL}$ and then performs the PDBL's calculation.
Despite introducing additional computation overhead (\ie, one $modmul$ operation) to get the correct inverse values, the benefits of the discard-then-recompute scheme far outweigh it.
This scheme fully utilizes the point data locality of the integrated PADD and PDBL calculation process to reduce data access, thus improving the performance of batch PMUL operations. 

Our design ensures constant time for the EC operations to realize side-channel protection~\cite{Attacks96}. To prevent any leakage of timing information related to the secret scalar in the PUML operation, instead of omitting unnecessary operations, the PMUL operation conducts a dummy operation with the point-at-infinity when required (for instance, zeros of the scalar are absorbed without revealing timing information, as shown in algorithm~\ref{alg:batch-pmul});

%% file: modular_design.tex
\section{Modular Arithmetic Design}

In this section, we first introduce the design of our modular arithmetic, which mainly includes a general $modmul$ algorithm for any modulus. Then we also develop curve-specific modular reduction optimization for the SM2 curve, which is the standard in China. 
% We carefully schedule and optimize the algorithm at the assembly level using PTX (Parallel Thread Execution) and ISA (Instruction Set Architecture) to achieve the theoretical optimal performanc at the micro architecture level.

% we propose an efficient and concise reduction arithmetic for modular addition (subtraction) and multiplication (square) of Curve25519/448, by carefully scheduling and optimizing the algorithm at assembly level using CUDA parallel thread execution (PTX) instruction set architecture (ISA) instructions.

\subsection{Modular Multiplication Optimization}
\label{subsec:mul}
\subsubsection{Identify bottlenecks in modular multiplication solutions.} 
In this section, we aim to identify the potential optimization of $modmul$. To this end, we benchmark the execution of SASS instructions at the microarchitectural level. SASS instructions directly affect the performance of the program, but the issuance and latency of SASS instructions are not openly documented. Official reports~\cite{gtc2019, cuda2024} only provide the throughput of native arithmetic operations per cycle per steaming processor. There is a huge gap between PTX instructions, with which developers program their computations, and SASS instructions, which are actually executed on GPUs. This gap is evident in two main areas. 1) How the arithmetic operations of PTX instructions are compiled into SASS instructions. The same arithmetic operations can be implemented with different PTX instructions by different programmers, such as different order of arithmetic operations, different storage allocation of the results of arithmetic operations, etc. Finally, the compiler also plays an important role in what SASS instructions are produced for an arithmetic operation. Therefore, it is critical to gain knowledge of the correspondence between arithmetic operations, PTX instructions as well as SASS instructions, including SASS Operators and register usage information.
2) Programs with the same arithmetic complexity in terms of PTX instructions may differ in their performance. Although we can get the total clock cycles and the total number of the different types of SASS instructions of a program, it is unclear how the SASS instructions are pipelined on the GPU, and hence we cannot estimate the performance of the program accurately. The pipelining may render a program with a larger number of SASS instructions run faster than another with a smaller number. This is because the instruction issue rates depend on the types of SASS instructions and their execution order. There are cases where the program with more SASS instructions may perform better due to its higher instruction issue rates. 
The lack of documentation drives us to simulate the pipeline of instruction issuance, which enables us to identify the bottleneck of a program and the potential for optimization. Therefore, we also need to benchmark the issue rates of the different types of SASS instructions. 
%After that, we define performance metric by multiplying instruction issue rate of the corresponding type of SASS instructions and the number of instructions.

%During the implementation, we use PTX instruction for programming, which is then converted into hardware instructions, namely SASS instruction, by the Nvidia compiler and finally mapped to the GPU for execution. Therefore, we must benchmark the execution of SASS instructions at the GPU microarchitecture level, because SASS instructions directly affect the performance of the program, but the issuance and latency of SASS instructions are not exposed to programmers. Official reports~\cite{gtc2019, cuda2024} only provide the throughput of native arithmetic operations per cycle per steaming processor, which is a huge gap between PTX instruction and SASS instruction. Our goal is to simulate the pipeline of instruction issuance by counting the number of SASS instructions in a program. By multiplying the issuance rate of the corresponding type of SASS code by the number of instructions to explore the bottleneck of the program and whether there is room for optimization.

We tested four types of SASS instructions related to computing, which represented a large proportion in the previous solution~\cite{gao2020dpf, gao2021dpf, emmart2018faster, dong2018sdpf, zheng2014exploiting, cgbn, dong2018towards, sppark2024}. Among them, DFMA and DADD denote the multiply-add and addition operations for two double-precision floating-point numbers, respectively. IMAD and IADD3 are instructions for multiply-add and addition operations on 32-bit integers, as shown in Table~\ref{tab:analysis-inst}. We count the total number of cycles required for a warp to execute 10,000 SASS instructions of the same type (with and without data dependency). By combining the results in the official reports with those of our benchmark, we can obtain an estimate of the issue rate per warp scheduler of SASS instructions, as shown in Table~\ref{tab:analysis-inst}. It is evident that both data-independent DFMA and IMAD instructions are launched at intervals of 4 cycles. 

% In addition, we also count the total number of cycles required for a warp to execute 10,000 SASS instructions of different types (without data dependencies).
% In addition, we conducted experiments to execute different types of SASS instructions simultaneously. 
\begin{table}[h]
    \vspace{-0.25cm}
    \centering
    \setlength{\abovecaptionskip}{0.1cm}
    \caption{GPU Ampere micro architecture level SASS instructions issuance analysis}
    \resizebox{
    \textwidth}{!}{
    \label{tab:analysis-inst}
    \begin{tabular}{ccccc}
        \toprule %[2pt]设置线宽
        \shortstack{arithmetic \\ operation} & \shortstack{PTX \\ instructions} & \shortstack{SASS \\ Operators} & \shortstack{dependent / independent \\ issue rate \\ (per warp scheduler)} & \shortstack{meaning} \\
        \midrule %[2pt]  
        \shortstack{floating-point \\ mul-add} & fma.rz.f64 d, a, b, c & DFMA & \shortstack{$\frac{1}{8}$ / $\frac{1}{4}$ cycle} & $d=a*b+c$ \\ 
        \midrule %[2pt]
        \shortstack{floating-point \\ add} & add.rz.f64 d, a, b& DADD & \shortstack{$\frac{1}{8}$ / $\frac{1}{4}$ cycle} & $d=a+b$  \\ 
        \midrule %[2pt]
        \shortstack{integer \\ mul-add} & \shortstack{madc.cc.lo.u32 $(d)_{lo}$, a,b,$(c)_{lo}$ \\ madc.cc.hi.u32 $(d)_{hi}$, a,b,$(c)_{hi}$} & \shortstack{IMAD.WIDE.X} & \shortstack{$\frac{1}{8}$ / $\frac{1}{4}$ cycle} & \shortstack{$cc, (d)_{lo}=(a*b)_{lo}+(c)_{lo}+cc$ \\ $(d)_{hi} = (a*b)_{hi}+(c)_{hi}+cc$}\\ 
        \midrule %[2pt]
        \shortstack{integer \\ add} & addc.cc.u32 d,a,b; & IADD3.X & \shortstack{$\frac{1}{4}$ / $\frac{1}{2}$ cycle} & $cc, d=a+b+cc$\\ 
        \bottomrule %[2pt] 
    % \end{tabular}
    \end{tabular}
    }
    % \begin{tablenotes}
    % \centering
    % \item * If .cc specified, carry-out written to predicate register.
    % \end{tablenotes}
    \vspace{-0.25cm}
\end{table}
In addition, we also conducted experiments involving a warp issuing 10,000 different types of SASS instructions (without data dependencies) consecutively and tallied the total number of cycles required for processing all the instructions. The 10,000 instructions consist of two SASS instructions (any two types shown in Table~\ref{tab:analysis-inst}), arranged in a 1:1 ratio. %and 1:2 ratio. 
For instance, an IMAD instruction is followed by a DFMA instruction. The results indicate that when a warp issues one DFMA followed by one IMAD instruction, the total cycles required are the sum of issuing DFMA or IMAD alone, which means that although these two types of instructions run in different compute units, they are still launched at intervals of 4 cycles. Similarly, the interval between the DFMA instruction and the DADD instruction, as well as the IMAD instruction and the DADD instruction, remains at 4 cycles. However, the interval between the IMAD, DFMA, or DADD instruction and the IADD3 instruction is 2 cycles. 

Theoretically, a 256-bit integer represented in base $D=32$, necessitates $m=8$ 32-bit integers for storage. As shown in Algorithm~\ref{alg:ori_montmul}, the $modmul$ operation of two 256-bit integers consists of two parts. The integer multiplication requires $m^2=64$ IMAD instructions, and the modular reduction requires $m+m^2=72$ IMAD instructions. This results in a total of 136 IMAD instructions. When considering $D=52$ in the solution (DPF) presented in~\cite{emmart2018faster}, storing a 256-bit integer requires $5$ double-precision floating-point numbers. Due to considerations regarding floating-point precision, storing the product of two 52-bit integers necessitates 2 DFMA instructions. So, in $modmul$ operation in this solution, integer multiplication demands $2*m^2=50$ DFMA instructions, while the modular reduction requires $2*m^2=100$ DFMA instructions and $m=5$ IMAD instructions. Moreover, the solution incorporates extra DADD instructions and data conversions to ensure precision.

We used the GPU performance analysis tool NVIDIA Nsight Compute~\cite{nsightcompute} to disassemble the fastest $modmul$ implementation~\cite{cgbn} based on integer instructions. This implementation uses 1, 2, and 4 threads with  $D=32$, and is referred to as CGBN-1, CGBN-2, and CGBN-4, respectively. Additionally, we disassembled a solution~\cite{emmart2018faster, gao2020dpf} based on a mixture of multi-precision floating-point instructions and integer arithmetic instructions, which uses 1 thread and has a parameter $D=52$. This solution is named DFP-1.

\begin{table}[ht]
% \begin{wraptable}{r}{0.6\linewidth}
    \vspace{-0.25cm}
    \centering
    \caption{Instructions analysis of a $modmul$ in different solution for SM2 curve}
    \label{tab:analysis-modmul}
    \begin{tabular}{ccccccc}
        \toprule %[2pt]设置线宽
         & total inst & DFMA & DADD & IMAD & IADD3 & SHFL \\ 
        % \midrule %[2pt]  
        \hline
        DPF-1 & 325 & \textbf{100} & \textbf{50} & 29 & 114 & 0 \\ 
        % \midrule %[2pt]
        \hline
        CGBN-1 & 227 & 0 & 0 & 185 & 32 & 0   \\ 
        % \midrule %[2pt]
        \hline
        CGBN-2 &  382 & 0 & 0 & 148 & 94 & \textbf{50} \\ 
        % \midrule %[2pt]
        \hline
        CGBN-4 &  584 & 0 & 0 & 188 & 144 & \textbf{100} \\ 
        \bottomrule %[2pt] 
    \end{tabular}
    \vspace{-0.25cm}
\end{table}
% \end{wraptable}

The results of the performance test depicted in Fig.~\ref{fig:performance_montmul} indicate the following performance ranking: CGBN-1 outperforms DPF-1, which is in turn more efficient than CGBN-2 and CGBN-4. As shown in Table~\ref{tab:analysis-modmul}, the primary performance bottlenecks for CGBN-4 and CGBN-2 arise from the SHFL instructions necessary for inter-thread communication~\cite{cuda2024}. These instructions take approximately $5$ clock cycles~\cite{wang2017communication} and introduce synchronization overhead within the $modmul$ operation. DPF-1 has attempted to replace some IMAD instructions with DFMA instructions. Regrettably, despite these two types of instruction operating in different computational units, they cannot be executed concurrently, as confirmed by our benchmark results~\ref{tab:analysis-inst}. Furthermore, additional DADD instructions and data conversion instructions have hindered DPF-1 from outperforming the integer-based implementation CGBN-1. Although CGBN-1 has shown impressive performance, a microarchitecture analysis reveals that actual usage of IMAD instructions is 38\% higher than theoretical predictions, suggesting that there is still potential for further optimization.

In general, the main challenge in speeding up $modmul$ lies in minimizing the number of IMAD instructions and reducing the dependencies between instructions. Additionally, it is important to avoid register bank conflicts, as these can also delay the issuance of instruction~\cite{jia2018dissecting, jia2021dissecting, gtc2019}.

\begin{wrapfigure}{r}{0.65\textwidth}
    \vspace{-0.5cm}
    \centering
    \setlength{\abovecaptionskip}{0.1cm}
    % \vspace{0.1cm}
    \includegraphics[width=0.65\textwidth]{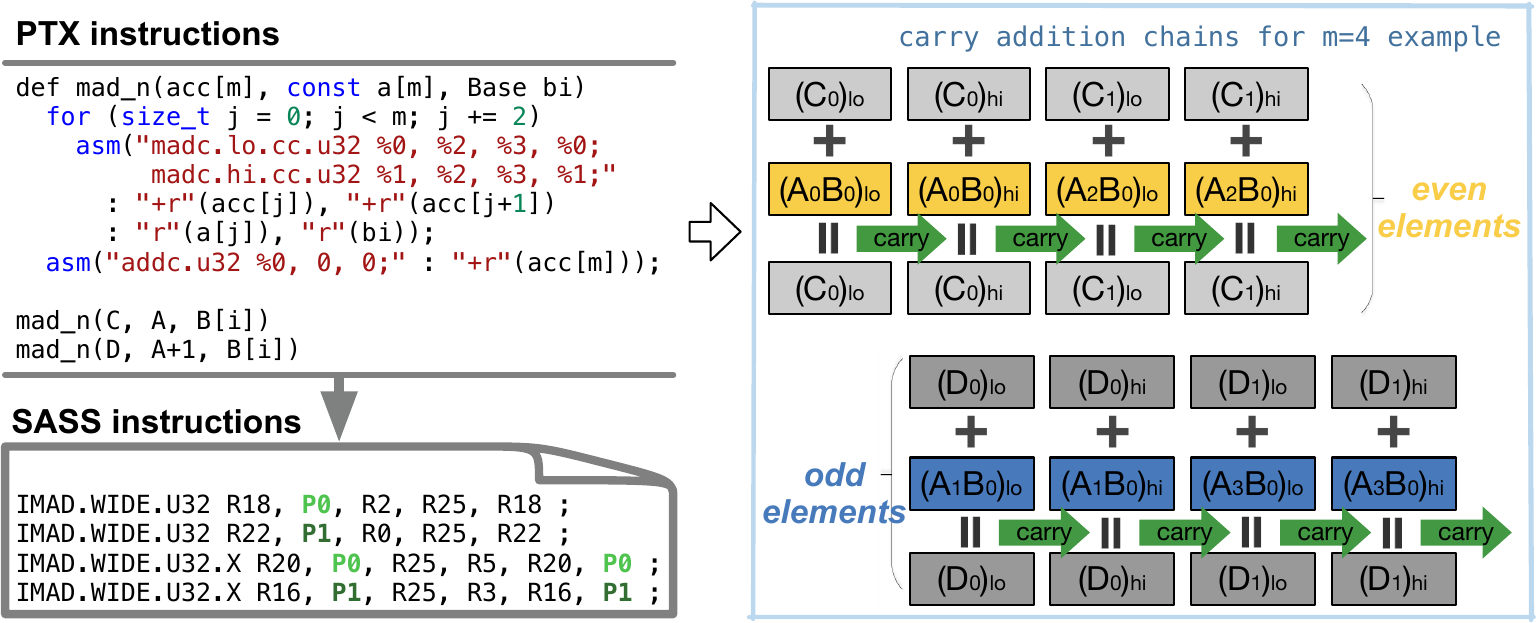}
    \caption{Divergent carry addition chains for odd and even word-size integer} 
    \label{fig:mad_n}
    \vspace{-0.25cm}
\end{wrapfigure}
\subsubsection{Minimize IMAD instructions.} %by chaining divergent carry additions for odd and even elements in multiplication.} ŵ
Based on the preceding analysis, as shown in Table~\ref{tab:analysis-inst}, the native arithmetic operations for implementing $modmul$ operation is the multiply-add arithmetic operation defined as $(\text{carry-out}, c) = a * b + c + \text{carry-in}$. Here, $a$ and $b$ are 32-bit integers, $c$ is a 64-bit integer consisting of a high 32-bit word and a low 32-bit word, and both carry-in and carry-out are each 1-bit. After compilation, the SASS instructions executed on the hardware is \verb|IMAD.WIDE.U32.X R1, P0, R2, R3, R1, P0;|. In these instructions, R1, R2, and R3 are general-purpose registers that store $c$, $a$, and $b$, respectively, while P0 is a predicate register used to pass the carry information to another SASS instruction. 
% In the case of 256-bit integer multiplication,
Improper propagation of carry information can lead to instruction overhead. For example, in the study of~\cite{dong2018towards}, if the high word carry and the low word carry of $a * b$ are propagated separately, the number of IMAD instructions can be doubled compared to the theoretical predictions.

Our implementation of multiplication carry propagation, akin to sppark~\cite{sppark2024}, effectively circumvents such issues. As depicted in Fig.~\ref{fig:mad_n}, the product of integers A and B can be computed by generating $m$ row-by-row multiplications with \verb|mad_n| function, specifically $A*B_i$, where $i$ ranges from $0$ to $m-1$. The carry from the high word of $A_j*B_i$ can be passed as the carry input to the low word of $A_{j+2}*B_i$, where $j$ ranges from $0$ to $m-1$. It includes the product of all even elements of A and $B_i$ and avoids the extra addition instructions caused by the conflict when the carry from the high word of $A_j*B_i$ carries over to the high word of $A_{j+1}*B_i$. This propagation of the carry can also apply to the remaining odd elements of $A$, such as the carry from the high word of $A_{j+1}*B_i$ can be passed as the carry input to the low word of $A_{j+3}*B_i$.

\subsubsection{Reorder IMAD instructions to reduce register moves and register bank conflicts.} 
% \begin{figure}
\begin{wrapfigure}{r}{0.65\textwidth}
    \centering
    \setlength{\abovecaptionskip}{0.1cm}
    \vspace{-0.5cm}
    \includegraphics[width=0.65\textwidth]{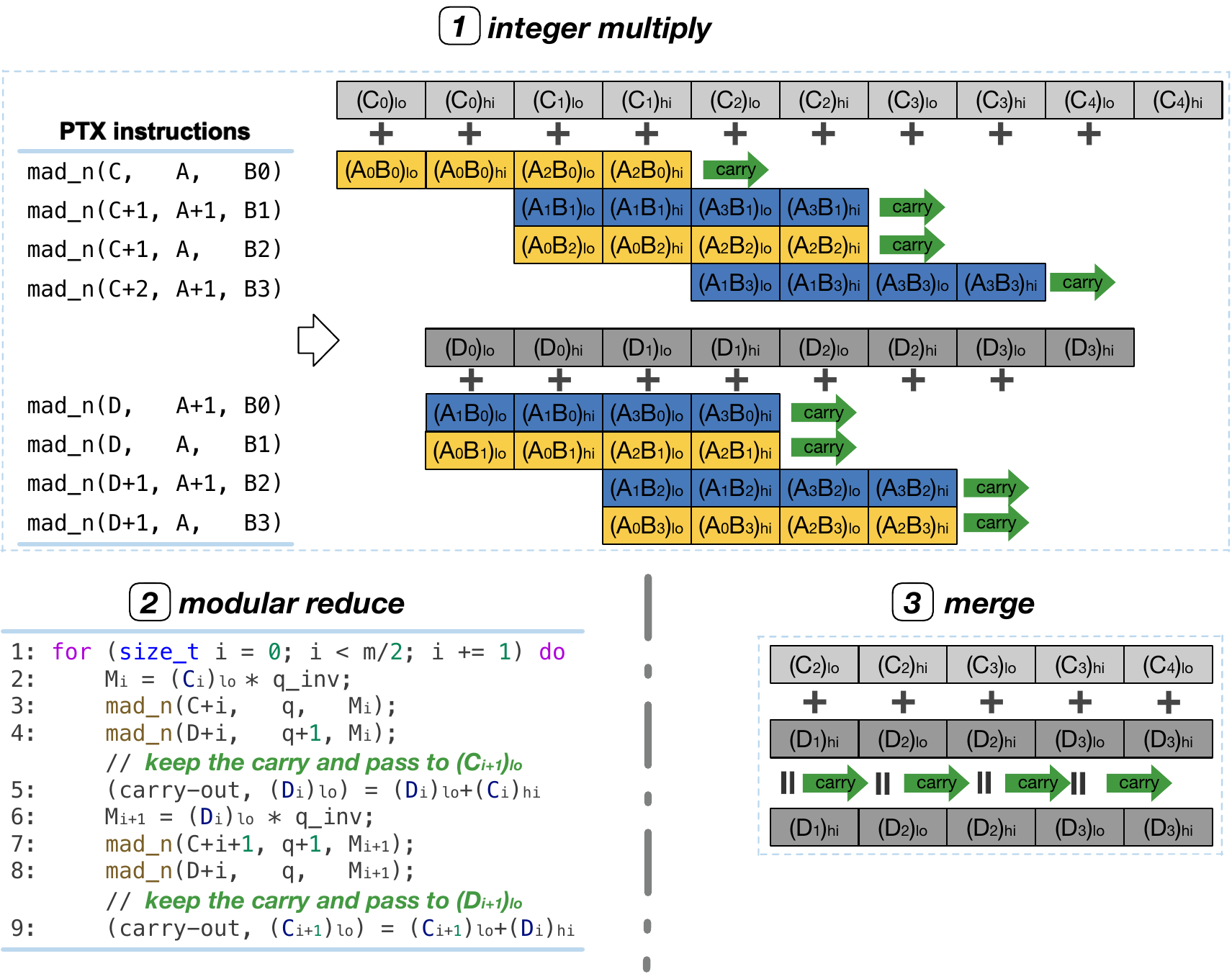}
    \caption{General $modmul$ algorithm for m=4} 
    \label{fig:sos}
    \vspace{-0.5cm}
\end{wrapfigure}
% \end{figure}
After customizing the one row-by-row multiplication with two \verb|mad_n| functions, we still need to pay attention to the IMAD order when applying it to the integer multiplication stage of the $modmul$ operation because register move and bank conflicts will also affect the instructions' issuance rate. 
% In the multiplication phase, 
when processing $m$ row-by-row multiplications in integer multiplication of $modmul$ operation, we modify the write-back registers of the IMAD instruction, as shown in Fig.~\ref{fig:sos}. 
% a departure from the approaches of sppark and CGBN. 
Specifically, the product results of the even elements in $A*B[0]$ are accumulated in the $C$ array, while those of even elements in $A*B[1]$ are accumulated in the $D$ array. This adjustment offers the advantage of eliminating the need to separately read the high word $(C_0)_{hi}$ of the $(C_0)$ register and the low word $(C_1)_{lo}$ of the $(C_1)$ register, thus reducing the movement and reading operations of the registers.

During the modular reduction phase of $modmul$ operation, as shown in Algorithm~\ref{alg:ori_montmul}, while accumulating the product of $q*M_i$ in each iteration, sppark ensures result accuracy by shifting the $C$ or $D$ array 64 bits to the right, such as $C_i=C_{i+1}+q*M_1$. However, the corresponding SASS instructions requires the IMAD instruction to access four registers, potentially causing a one-cycle delay in instruction issuance due to register bank conflicts~\cite{jia2018dissecting, jia2021dissecting, gtc2019}. We employ an in-place write-back strategy, namely $C_i=C_i+q*M_1$, which allows the IMAD instruction to access three registers simultaneously. To ensure that the results of the $C$ and $D$ arrays are shifted 64 bits to the right in each iteration without increasing the number of IMAD instructions like CGBN, we utilize predicate registers to store carries as shown in phase (2) in Fig.~\ref{fig:sos}. For instance, when $i=0$, the value of $(C_0)_{hi}$ is added to $(D_0)_{lo}$, and the \verb|carry-out| is stored in the predicate register to be passed to $(C_1)_{lo}$ in the subsequent \verb|mad_n| function.
Subsequently, the IADD3 instruction is employed to accumulate the $C$ and $D$ array to produce the final result in the merge phase in Fig.~\ref{fig:sos}. Upon disassembly of our design, the IMAD instruction in a $modmul$ operation aligns with the theoretical prediction.

\subsection{Modular Reduction Optimization on SM2 Curves}
\label{subsec:reduce}
To further reduce the number of IMAD instructions, we observed that, for the SM2 curve, the IADD3 instruction with a higher issuance rate can be used to replace the number of IMAD instructions. The unique forms of the SCA-256 prime modulus on the SM2 curve facilitate a distinct modular reduction phase. On the SM2 curve, $q = 2^{256} - 2^{224} - 2^{96} + 2^{64} - 1$, and therefore $q\_inv$ of $q$ in Algorithm~\ref{alg:ori_montmul} is equal to 1. A previous work~\cite{hu2019high}, an ASIC-based accelerator for the SM2 curve, also utilizes this property. However, its design is not portable to GPU as it introduces a multitude of intermediate results that require more registers and introduces a lot of addition operation. 

% \begin{figure}[h]
%     \centering
%     \setlength{\abovecaptionskip}{-0.08cm}
%     \vspace{-0.4cm}
%     \includegraphics[width=0.8\linewidth]{figure/reduce1.pdf}
%     \caption{modular reduction optimization for lines 2 to 4 when $i=0$}
%     \label{fig:reduce1}
%     \vspace{-0.5cm}
% \end{figure}
So in the middle part of Fig.~\ref{fig:sos}, $q\_inv = 1$ in line 2 and line 6 is specified, rendering $M_i$ equivalent to $(C_i)_{lo}$ or $(D_i)_{lo}$. let us incorporate the value of $q$ into the reduction phase, the \verb|mad_n| in the line 3, 4, 6 and 7 can be simplified to addition instead multiplication operation. For example, at the iteration $i=0$, lines 2 to 4 can be simplified using Eqn.~\ref{form:reduce1}. Therefore, 8 subtractions are required to construct $q*(C_0)_{lo}$, and 10 additions are required to update the $C$ array (from $(C_0)_{lo}$ to $(C_4)_{hi}$).
\begin{equation}
\label{form:reduce1}
    \begin{aligned}
        C&=C+(C_0)_{lo} * q=\sum_{i=0}^7((C_i)_{lo}+(C_i)_{hi}*2^{32})*2^{i*32} + (2^{256} - 2^{224} - 2^{96} + 2^{64} - 1)*(C_0)_{lo}\\
        &=(C_0)_{hi}*2^{32}+((C_1)_{lo}+(C_0)_{lo})*2^{64}+((C_1)_{hi}-(C_0)_{lo})*2^{96}+(C_2)_{lo}*2^{128}+(C_2)_{hi}*2^{160}+(C_3)_{lo}*2^{192}\\&+((C_3)_{hi}-(C_0)_{lo})*2^{224}+((C_4)_{lo}+(C_0)_{lo})*2^{256}+(C_4)_{hi}*2^{288}+\sum_{i=5}^7((C_i)_{lo}+(C_i)_{hi}*2^{32})*2^{i*32}
    \end{aligned}
    % \vspace{-0.6cm}
\end{equation}

After that, both arrays $D$ and $(C_0)_{hi}$, required by line 5, remain unchanged, allowing lines 6 to 8 to be simplified in a similar manner. Consequently, in each iteration of the reduction phase, $8*2$ subtractions are still necessary to construct $(C_i)_{lo} * q$ and $(D_i)_{lo} * q$, and $10 * 2$ additions are required to update the arrays $C$ and $D$. 

Nonetheless, there is potential for further optimization. As depicted in Fig.~\ref{fig:sos}, the merge phase of $modmul$ does not necessitate precise values for $C_0, C_1$ and $D_0, D_1$, but merely requires accurate carry information. Hence, we can simplify the computation from line 2 to line 8 when $i=0$ using Eqn.~\ref{form:reduce2}. In the calculation of $tmp$, we employ one addition operation and use the predicate register to store the carry information. When computing $(C_i)_{lo} * q + ((C_i)_{hi} + (D_i)_{lo}) * q$, we require an additional eight 32-bit integers and seven subtractions with IADD3 instructions, as we can disregard the specific value of the lower 64-bits. When updating the array $C'$, we also ignore the lower 64 bits and simply add the carry information stored in the predicate register, which can be accomplished through nine additions with IADD3 instructions. Compared to the previous work in~\cite{hu2019high}, which requires 176 addition and subtraction instructions, we only need $(1+7+9+1)*4=72$ addition and subtraction instructions to complete the reduction phase.
\begin{equation}
\label{form:reduce2}
    % \vspace{-0.5cm}
    \begin{aligned}
        (&cc,tmp)=(C_0)_{hi}+(D_0)_{lo}\\
        C&=C+(C_0)_{lo} * q+((C_0)_{hi}+(D_0)_{lo})*q\\
        C'&=(cc+(C_1)_{lo}+(C_0)_{lo})*2^{64}+((C_1)_{hi}-(C_0)_{lo}+tmp)*2^{96}+((C_2)_{lo}-tmp)*2^{128}+(C_2)_{hi}*2^{160}\\&+(C_3)_{lo}*2^{192}+((C_3)_{hi}-(C_0)_{lo})*2^{224}+((C_4)_{lo}+(C_0)_{lo}-tmp)*2^{256}+((C_4)_{hi}+tmp)*2^{288}\\&+\sum_{i=5}^7((C_i)_{lo}+(C_i)_{hi}*2^{32})*2^{i*32}
    \end{aligned}
    % \vspace{-0.5cm}
\end{equation}

% \begin{figure}[h]
%     \centering
%     \vspace{-0.4cm}
%     \setlength{\abovecaptionskip}{-0.08cm}
%     \includegraphics[width=0.8\linewidth]{figure/reduce3.pdf}
%     \caption{modular reduction optimization for lines 2 to 8 when $i=0$}
%     \label{fig:reduce2}
%     \vspace{-0.5cm}
% \end{figure}

%% file: evaluation.tex
\section{Experimental Evaluation}
\label{sec:eval}
%In this section, we present and discuss the evaluation results as follows.

\subsection{Methodology}
\para{\textit{Experimental Setup:}}
Our experiments are conducted on a virtual machine with an Intel Xeon processor (Skylake, IBRS) and \SI{72}{GB} DRAM, running Ubuntu 18.04.2 with CUDA Toolkit version 12.2.
It is equipped with one NVIDIA Ampere A100 GPU (each with \SI{40}{GB} memory) and one NVIDIA Votal V100 GPU (each with \SI{32}{GB}). Because the optimization of EC operations in \oursys{} relies on the new feature of L2 cache persistence in A100 GPU, we only evaluate the throughput performance of \ecc{} algorithm (Section~\ref{subsec:eval_ecdsa}), EC operations (Section~\ref{subsec:eval_pmul}) and application (Section~\ref{subsec:eval_app}) on A100 GPU. Because the optimization of $modmul$ operation in \oursys{} is general and does not rely on this feature, we evaluate it on both V100 and A100 GPUs (Section~\ref{subsec:eval_montmul}). We use random synthetic data generated by RapidEC based on the SM2 curve to evaluate the performance of \ecc{} algorithm 
% (Section~\ref{subsec:eval_ecdsa})
, EC operations 
% (Section~\ref{subsec:eval_pmul})
, and $modmul$ operation 
% (Section~\ref{subsec:eval_montmul})
performance. And we used real workload data to evaluate the application 
% (Section~\ref{subsec:eval_app}) 
performance. 

% \para{\textit{Workloads:} }

\para{\textit{Baselines:}}
We compare \oursys{} with the state-of-the-art method for GPU-accelerated \ec{} operations, called RapidEC~\cite{feng2022accelerating}.

\para{\textit{Performance Metrics:}}
We focus on enhancing the throughput of the GPU programs. Therefore, for the ECDSA algorithm, we use the number of signatures generated/verified per second (\ie{, signature/s}) as the performance metric for the signature generation and verification phases, respectively. 
The ECDH algorithm comprises two FPMUL operations and a single key value comparison. Given that the time taken for the latter is negligible, the throughput of the ECDH algorithm is dominated by the throughput of the FPMUL operation. Therefore, we only report the throughput value of the FPMUL operation.
Specifically, we record the calculation time $t$ when performing a batch of $n$ signature generations, verifications, and FPMUL operations, respectively, and calculate throughput as $\frac{n}{t}$, which is consistent with RapidEC.
%The data transfer between CPU and GPU can overlap well with the calculation of different batches, so we do not provide this part of the time separately.

\subsection{ECDSA Algorithm Performance}
\label{subsec:eval_ecdsa}
We evaluate the end-to-end throughput of signature generation and verification based on the SM2 curve, respectively.

\begin{table}[h]
% \begin{wraptable}{r}{0.35\textwidth}
    \vspace{-0.25cm}
    \centering
    \setlength{\abovecaptionskip}{0.1cm}
    \caption{Throughput Performance of ECDSA on NVIDIA A100 based on SM2 curve}
    \label{tab:overall-perf}
    \begin{tabular}{ccc}
        \toprule %[2pt]
        \shortstack{\textbf{Solution}} & \shortstack{\textbf{Signature} \\ \textbf{Generation}} & \shortstack{\textbf{Signature} \\ \textbf{Verification}} \\
        \midrule %[2pt] 
        RapidEC & 3,386,544 & 773,481 \\
        \oursys{} & 14,141,978 & 4,372,853 \\
        \bottomrule %[2pt]  
    \end{tabular}
    \vspace{-0.25cm}
\end{table}
% \end{wraptable}
\para{Signature Generation.}
We examine the throughput of signature generation in the ECDSA algorithm.
%We use random synthetic data generated by RapidEC based on the SM2 curve. 
% It's worth noting that the scalar and point values for different signature generations or verifications are the same in the RapidEC's benchmark, which prevents the issue of GPU warp divergences when computing batch PMUL operations concurrently. 
% Note that the SM2 curve uses 256-bit integers.
As shown in Table~\ref{tab:overall-perf}, the throughput of signature generation in \oursys{} is 14,141,978 signature/s, which achieves 4.18$\times$ speedup compared to the corresponding part of RapidEC.
The end-to-end performance advantage of \oursys{} can be attributed to its comprehensive optimization from the low-level $modmul$ to cryptographic operation. %, resulting in a significant improvement in end-to-end performance.

\begin{wrapfigure}{ r }{ 0.47 \textwidth } 
    \vspace{-0.25cm}
    \setlength{\abovecaptionskip}{-0.8cm}
    \centering
    % \vspace{0.1cm}
    \includegraphics[width=0.47\textwidth]{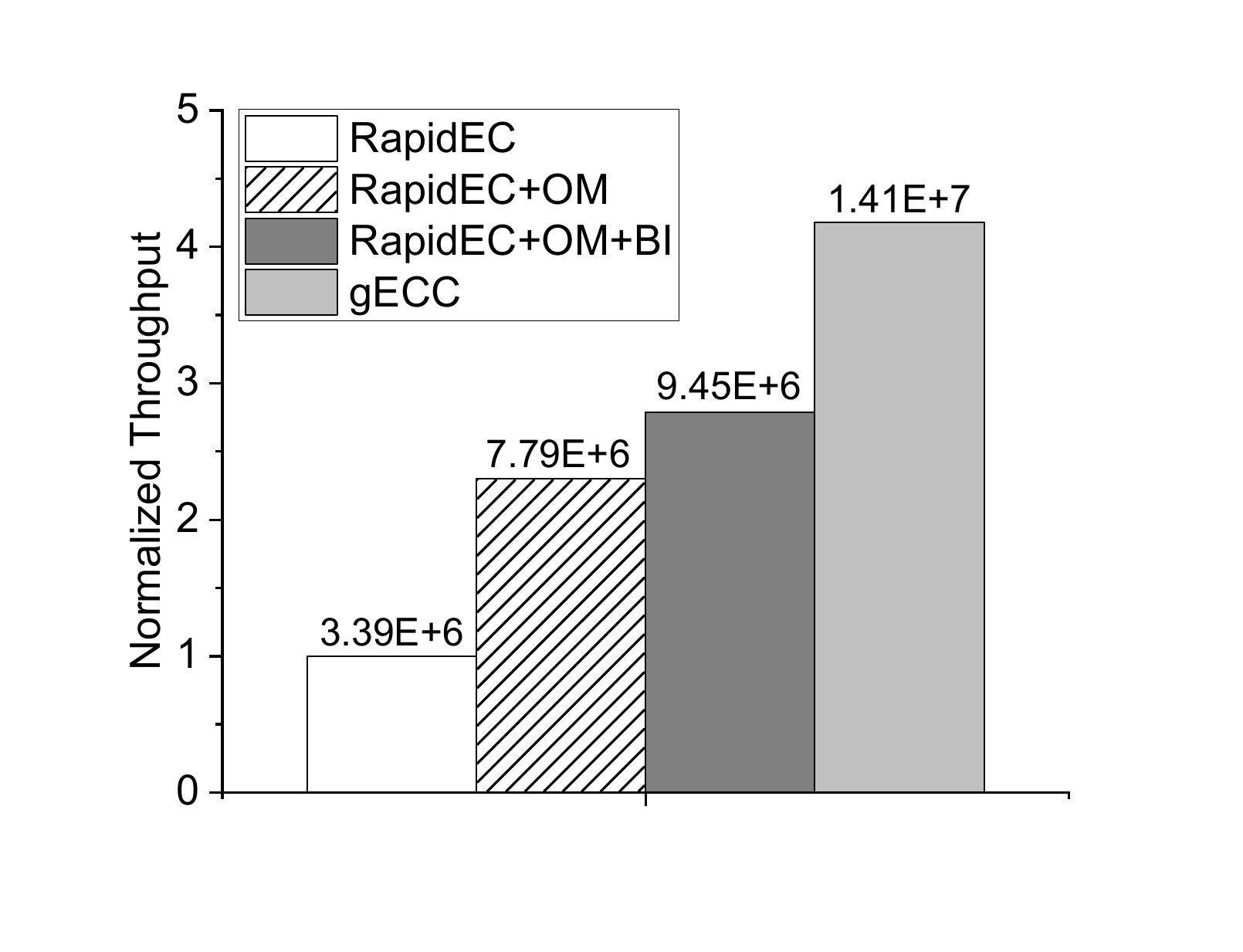}
    \caption{Breakdown analysis of signature generation} 
    \label{fig:sg-bkana}
    \vspace{-0.25cm}
\end{wrapfigure}
To take a closer look, we investigate the improvement brought about by the major optimization techniques of \oursys{} for the generation of signatures.
This generation mainly contains a $modinv$ operation and an FPMUL operation.
First, to verify the improvement brought about by our optimized modular arithmetic, we re-implemented the RapidEC solution with our optimized modular arithmetic (``RapidEC+OM'').
Then, we evaluate an alternative solution (``RapidEC+OM+BI''), which further incorporates batch modular inversion optimization presented in Section ~\ref{subsec:pminv}.
Finally, \oursys{} (``\oursys'') contains all our proposed optimizations.
It adopts the optimized FPMUL by leveraging the memory management optimization (Section~\ref{subsec:batch-padd}). As shown in Fig.~\ref{fig:sg-bkana}, the breakdown analysis shows that RapidEC+OM achieves $2.3\times$ speedup against RapidEC, 
and RapidEC+OM+BI further improves the performance by $21\%$ over RapidEC+OM.
\oursys can achieve another $50\%$ improvement against RapidEC+OM+BI.
This verifies that batch modular inversion based on Montgomery's Trick could significantly reduce the number of $modinv$ operations, and the GAS mechanism could minimize the overall latency and fully utilize the high degree of parallelism provided by GPU.
% Furthermore, the batch PADD operations based on affine coordinates reduce the overall computation overhead, and the multi-level cache management optimization also reduces the memory access overhead. 
Furthermore, the batch PADD operations based on affine coordinates reduce the overall computation overhead.
\begin{wrapfigure}{ r }{ 0.47 \textwidth } 
    \vspace{-0.25cm}
    \setlength{\abovecaptionskip}{-0.8cm}
    \centering
    % \vspace{0.1cm}
    \includegraphics[width=0.47\textwidth]{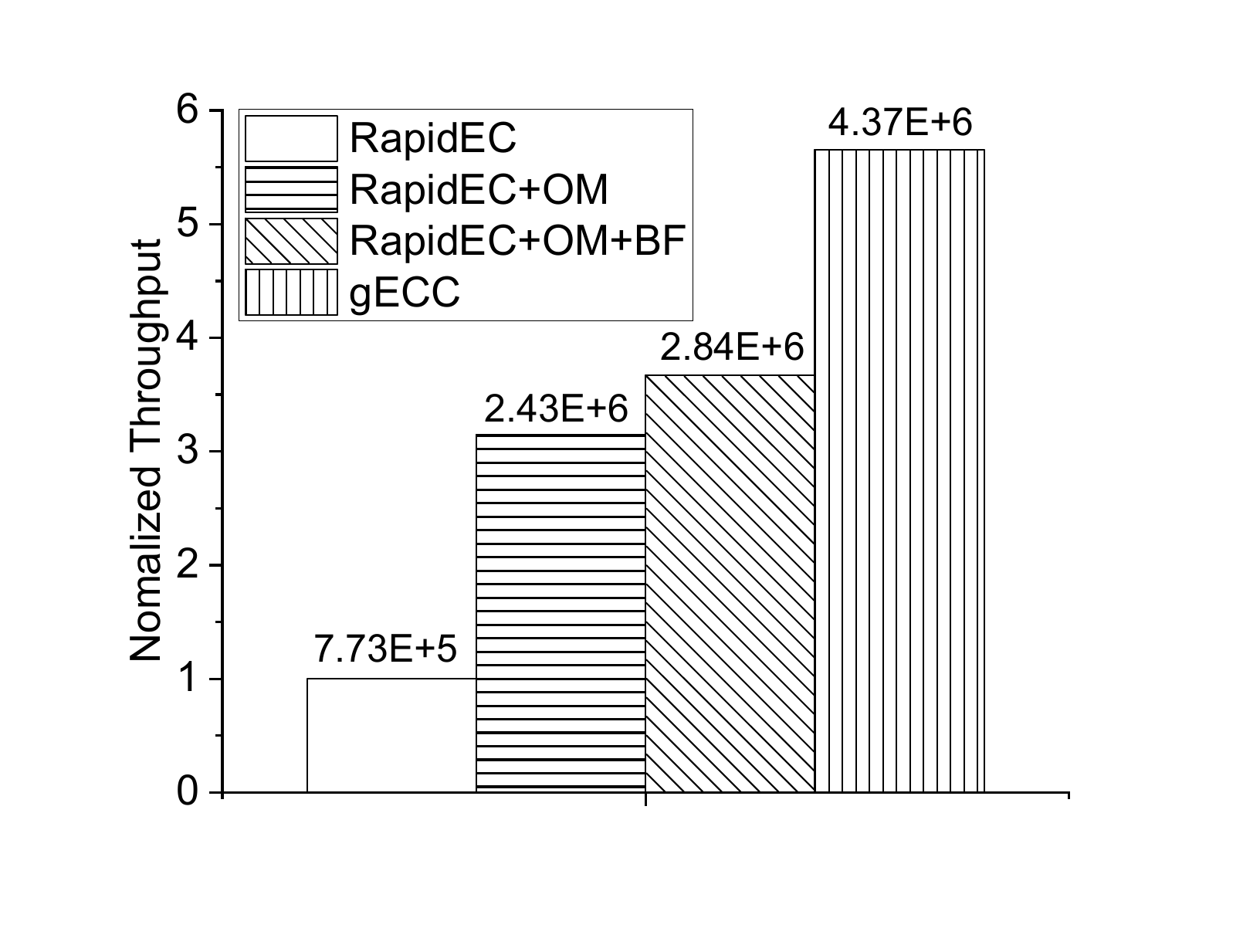}
    \caption{Breakdown analysis of signature verification} 
    \label{fig:sv-bkana}
    \vspace{-0.25cm}
\end{wrapfigure}
\para{Signature Verification.}
In this experiment, we examine the throughput of signature verification of the ECDSA algorithm, which mainly contains one FPMUL operation and one UPMUL operation. 
% Similar to the previous experiment, we conducted the experiments on random synthetic data created by RapidEC based on the SM2 curve.
As shown in Table~\ref{tab:overall-perf}, \oursys{}'s throughput of signature verification is 4,372,853 signature/s, which achieves 5.65$\times$ speedup compared to RapidEC.

We provide a breakdown analysis of the signature verification process as follows. 
First, we re-implemented the RapidEC solution with our optimized modular arithmetic (``RapidEC+OM''). 
% Then, we experiment with a variant of \oursys{}, \oursys{}$-$BF, a \oursys{} variant without accelerating the UPMUL operation with the optimization proposed in Section~\ref{sec:pminv},
Then, we evaluate an alternative solution (``RapidEC+OM+BF'') that also accelerates the FPMUL operation with the optimization proposed in Section~\ref{subsec:pminv}.
% Then, we evaluate two intermediate solutions (``BG-SG w. FPMUL'' and ``\oursys{-SV w. no-mem-opt}'').
% The first one is that only accelerates the fixed-point multiplication with the kernel-fusion optimization, and the other accelerates both kind of PMUL operations with proposed optimization in Section~\ref{subsec:batch-pm}.
Finally, ``\oursys'' contains all our proposed optimizations, including data-locality-aware kernel fusion (Section~\ref{subsec:batch-pmul}) and memory management (Section~\ref{subsec:batch-padd}) optimizations.
% for both types of PMUL operations.
As shown in Fig.~\ref{fig:sv-bkana}, the breakdown analysis shows that RapidEC+OM achieves $3.2\times$ speedup against RapidEC.
Signature verification contains more calculations than signature generation, so our optimized modular arithmetic could provide a more significant speedup compared to RapidEC.
With the accelerated FPMUL operation, RapidEC+OM+BF further improves the throughput by $17\%$ over RapidEC+OM.
By accelerating the UPMUL operation, 
% the most time-consuming part of signature verification,
\oursys achieves another $54\%$ improvement over RapidEC+OM+BF.

\subsection{Performance Analysis of Batch PMUL on GPU}
\label{subsec:eval_pmul}
This set of experiments evaluates the throughput of PMUL based on the SM2 curve.

% \begin{wraptable}{ r }{ 0.65 \textwidth } 
\begin{table}[h]
    % \vspace{-0.25cm}
    \centering
    \setlength{\abovecaptionskip}{0cm}
    \caption{Throughput Performance of PMUL on NVIDIA A100 GPU based on SM2 curve}
    \label{tab:pm-perf}
    \setlength{\tabcolsep}{1.3mm}{
    \begin{tabular}{|c|c|c|c|c|}
    \hline
    \multirow{2}{*}{\shortstack{Batch size \\ per GPU SP}} & \multicolumn{2}{c|}{FPMUL operation per second} & \multicolumn{2}{c|}{UPMUL operation per second} \cr \cline{2-5}
     & RapidEC & \oursys{} & RapidEC & \oursys{} \\ \hline
    $2^{11}$ & 3,928,144 & 11,723,016 ($2.98\times$) & 1,476,365 & 6,270,010 ($4.25\times$) \\ \hline
    $2^{12}$ & 3,935,789 & 13,005,468 ($3.30\times$) & 1,405,325 & 6,561,449 ($4.67\times$) \\ \hline
    $2^{13}$ & 3,939,345 & 13,755,865 ($3.49\times$) & 1,427,610 & 6,689,186 ($4.69\times$) \\ \hline
    $2^{14}$ & 3,810,161 & 14,389,168 ($3.78\times$) & 1,464,398 & 6,766,892 ($4.62\times$) \\ \hline
    $2^{15}$ & 3,863,144 & 14,385,871 ($3.72\times$) & 1,471,910 & 6,837,619 ($4.65\times$) \\ \hline
    $2^{16}$ & 3,906,517 & 14,299,320 ($3.66\times$) & 1,477,929 & 6,732,287 ($4.56\times$) \\ \hline
    \end{tabular}
    }
    % \vspace{-0.25cm}
\end{table}
% \end{wraptable}
\para{Fixed Point Multiplication (FPMUL).}
First, we evaluate the throughput of the FPMUL operation. 
As shown in Table~\ref{tab:pm-perf}, \oursys{} achieves an average speedup of $3.16\times$ compared to RapidEC. 
To ensure GPU streaming processors load balancing, we set the batch size by the product of the number of GPU stream processors (432) and the amount of data processed by each processor.
As the batch size increases, the throughput of FPMUL operations in \oursys{} gradually increases until it saturates the GPU.
This is because, with a larger batch size, Montgomery's trick can bring about a greater degree of reduction in the computational complexity. 
As shown in Fig.~\ref{fig:cgpw}, the total duration of FPMUL computation in \oursys{} comprises three parts: compress step, GAS step, and DCWPA step.
The duration of the compress and DCWPA steps is linearly proportional to the batch size, while the duration of the GAS step is almost constant regardless of the batch size (Section~\ref{subsec:batch-inv}). 
Therefore, the throughput can increase as the batch size increases. When the batch size reaches about $432*2^{15}$, the duration of the GAS part only occupies a small portion of the total elapsed time, so the overall performance tends to be stabilized as the batch size continues to increase. RapidEC's throughput remains almost constant with different batch sizes, as it does not perform any optimization for batch processing. 
% This occurs because the increased memory access in \oursys{} can lead to computation stalls, resulting in incomplete utilization of GPU computing resources. Consequently, as the workload rises, the GPU Streaming Processor is able to schedule more waiting computational tasks until the computing resources reach their saturation point.
% \begin{comment}
\begin{figure}[ht]
    \vspace{-0.5cm}
    \setlength{\abovecaptionskip}{0.01cm}
    \label{fig:pmul}
    \subfigure[Breakdown analysis of fixed point multiplication]{
        \includegraphics[width=0.47\linewidth]{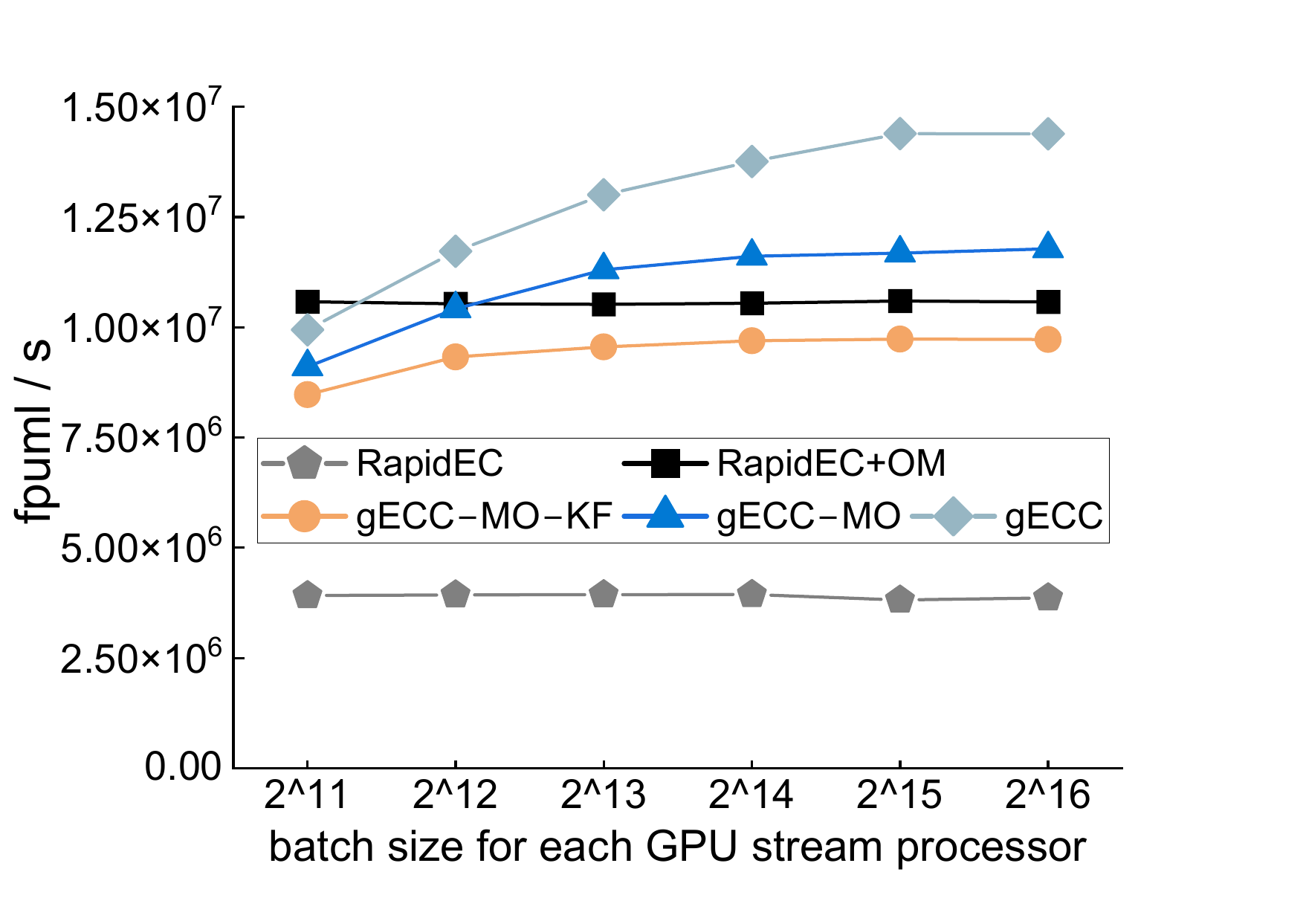}
        \label{fig:fpmul-bkana}
    }
    \subfigure[Breakdown analysis of unknown point multiplication]{
        \includegraphics[width=0.47\linewidth]{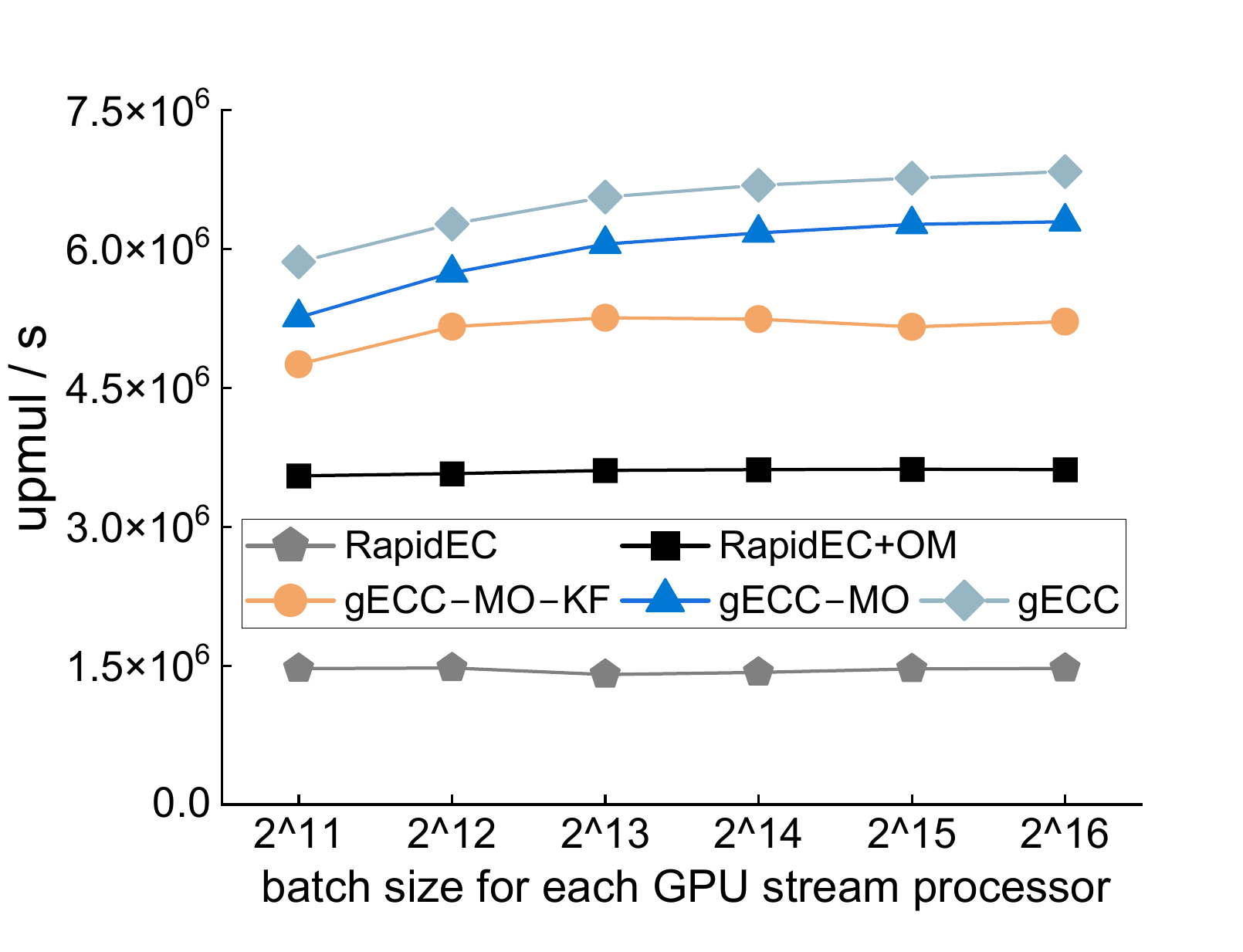}
        \label{fig:upmul-bkana}
    }
    % \Description{xxx}
    \caption{Breakdown analysis of Point Multiplication}
    \vspace{-0.5cm}
\end{figure}
% \end{comment}

In the next experiment, we investigate the impact of the major optimization techniques of \oursys{} on the FPMUL operation.
First, we re-implemented the RapidEC's FPMUL solution with our optimized modular arithmetic (``RapidEC$+$OM''). 
% using the double-and-add algorithm presented in~\ref{alg:danda} with preprocessed point data-based on the modular arithmetic in \oursys{}.
% Since the point $P$ is fixed, we can preprocess $2*P, 4*P, 8*P, ..., 2^{l-1}*P$ in advance to eliminate the PDBL operation for each bit of scalar in the algorithm~\ref{alg:danda}.
As shown in Fig.~\ref{fig:fpmul-bkana}, the throughput of RapidEC+OM is stable over different scales, which is up to 10,592,834 peration/s.
%The algorithm in Baseline generates much less intermediate data than \oursys{}, and all intermediate data of each FPMUL operation could be stored in registers, so the baseline can fully utilize the computing resources of GPUs at a smaller batch size.
We experiment with two variants of \oursys{}, \oursys{}$-$MO, a \oursys{} variant without the memory management mechanism, and \oursys{}$-$MO$-$KF, a variant without both the memory management and the data-locality-aware kernel fusion mechanisms.
The stabilized throughput of \oursys{}$-$MO$-$KF is lower than RapidEC+OM by $8\%$. 
The reason is that \oursys{}$-$MO$-$KF incurs a high overhead of global memory access due to the lack of relevant optimizations.
The throughput of \oursys{}$-$MO is about $10\%$ higher than that of RapidEC+OM because the kernel fusion optimization reduces the global memory access frequency and enhances data locality to a certain degree, thus improving the throughput of batch PADD operations.
% \begin{wrapfigure}{ r }{ 0.5 \textwidth } 
%     \vspace{-1cm}
%     \setlength{\abovecaptionskip}{-0cm}
%     \centering
%     % \vspace{0.1cm}
%     \includegraphics[width=0.5\textwidth]{gECC-FPMUL-BK-1210.pdf}
%     \caption{Breakdown analysis of fixed point multiplication} 
%     \label{fig:fpmul-bkana}
%     \vspace{-0.8cm}
% \end{wrapfigure}
Finally, the throughput of \oursys{} with all the optimizations can achieve $36\%$ and $23\%$ improvement over RapidEC+OM and \oursys{}$-$MO, respectively.  %is 14,389,168 opertion/s when the batch size is $439*2^{14}$.
This shows that our memory management optimization effectively optimizes global memory access through column-majored data layout and reduces data access latency through multi-level caching.

\para{Unknown Point Multiplication (UPMUL).}
This experiment examines the throughput of the UPMUL operation, the most time-consuming operation in \ecc{}.
As shown in Table~\ref{tab:pm-perf}, \oursys{} achieves, on average, $4.36\times$ speedup over the RapidEC.
RapidEC is implemented using the NAF (no-adjacent form) algorithm~\cite{hankerson2006guide}, which converts a scalar into its NAF representation with signed bits to reduce the number of PADD operations.
However, when multiple UPMUL operations are performed concurrently in a GPU Streaming Processor, it leads to severe warp divergence issues due to the different NAF values, resulting in higher inter-thread synchronization costs than \oursys{}.
% On the other hand, compared to the FPMUL operation, the throughput of UPMUL operation almost remains at a stable level as the batch size grows.
% The reason is that there is more computation in the UPMUL operation than in the FPMUL operation.
% So the batch UPMUL operations could fully utilize the GPU computing resources even with a smaller batch size.

% \begin{wrapfigure}{ r }{ 0.5 \textwidth } 
%     \setlength{\abovecaptionskip}{-0.1cm}
%     \centering
%     \vspace{-0.8cm}
%     \includegraphics[width=0.5\textwidth]{gECC-UPMUL-BK-1210.pdf}
%     \caption{Breakdown analysis of unknown point multiplication} 
%     \label{fig:upmul-bkana}
%     \vspace{-0.5cm}
% \end{wrapfigure}
We further investigate the improvement brought about by the optimizations of \oursys{} for the UPMUL operation.
Similarly, we re-implemented RapidEC's UPMUL operation using the NAF algorithm with our optimized modular arithmetic (``RapidEC+OM''). We also experiment with the same two variants of \oursys{}, \oursys{}$-$MO and \oursys{}$-$MO$-$KF. As shown in Fig.~\ref{fig:upmul-bkana}, the stabilized throughput of \oursys{}$-$MO$-$KF is $45\%$ higher than that of RapidEC+OM.
The reason is that the benefit of decreasing the computational complexity brought about by Montgomery's trick dominates the increased overhead of global memory access.
\oursys{}$-$MO brings about $20\%$ improvement and the memory management achieves a further $9\%$ improvement.
%The efficient kernel-fusion method with the recompute strategy fully utilizes the data-locality feature to reduces memory access and usage, improving the performance of batch PMUL.
%Moreover, \oursys explores opportunities for data access and computing overlap to reduce memory access latency.
%The multi-level cache management allocates appropriate cache according to the computing characteristics of different stages, ensuring efficient execution of batch EC operations.

\subsection{Performance Analysis of Modular Multiplication on the GPU}
\label{subsec:eval_montmul}
\begin{wrapfigure}{ r }{ 0.45 \textwidth } 
    \vspace{-0.25cm}
    \setlength{\abovecaptionskip}{-0.5cm}
    \centering
    % \vspace{0.1cm}
    \includegraphics[width=0.45\textwidth]{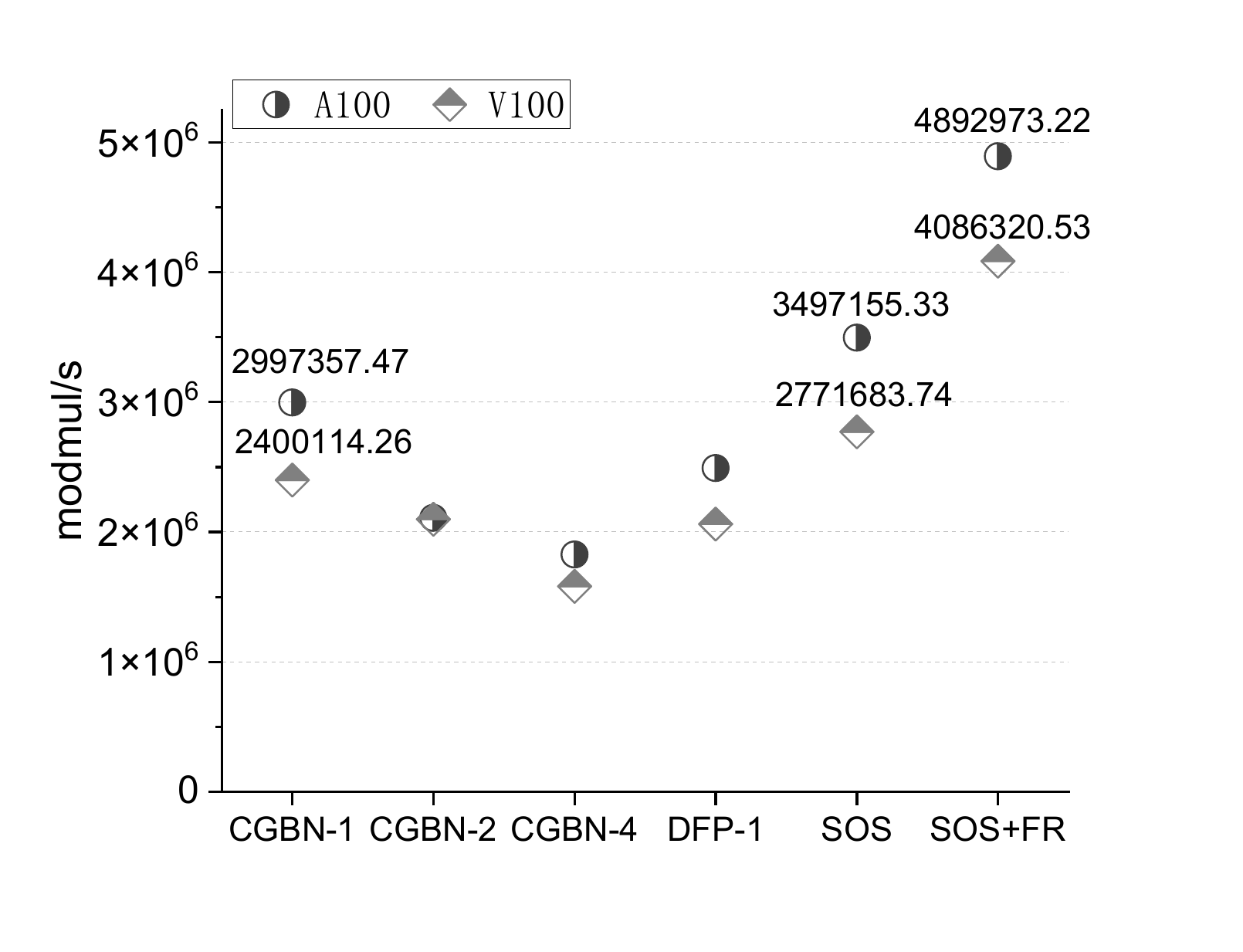}
    \caption{Throughput analysis of $modmul$ operation} 
    \label{fig:performance_montmul}
    \vspace{-0.25cm}
\end{wrapfigure}
In this section, we examine the throughput of the $modmul$ operation, the major underlying computation within \ecc{}. 
We compare against several accelerated $modmul$ implementations. CGBN~\cite{cgbn} is the finite field arithmetic library used by RapidEC. RapidEC uses four threads to calculate a $modmul$ operation, which we refer to as CGBN-4. 
In addition, we added two cases, using one or two threads to calculate a $modmul$ operation, referred to as CGBN-1 and CGBN-2, respectively. 
We also studied the solution presented in~\cite{gao2020dpf}, which is based on floating-point instructions and uses one thread to calculate a $modmul$ operation, which is referred to as DFP-1. 
First, we implement general $modmul$ operation ("SOS") as shown in Fig.~\ref{fig:sos}, which minimizing IMAD instructions and organizing the order of IMAD instructions (Section~\ref{subsec:mul}). Same as CGBN-1, SOS does not make special modular reduction optimizations for SCA-256 prime modulus and is a standard $modmul$ implementation which suitable for any prime moduli. Then, the full \oursys{} ("\oursys{}") adopts fast reduction for SCA-256 prime modulus (Section~\ref{subsec:reduce}).

As shown in Fig.~\ref{fig:performance_montmul}, the \oursys{} on SCA-256 prime modulus achieves $1.63 \times$ and $1.72 \times$ speedup against CGBN-1, and $2.68 \times$ and $2.58 \times$ speedups against CGBN-4 on the V100 and A100 GPU, respectively. 
To take a deeper look, we investigate the improvement brought by major \oursys{} proposals for $modmul$ operation. First, both SOS, \oursys{} and CGBN-1 avoid the inter-thread communication overhead in the CGBN-4, and significantly improve the throughput performance. SOS achieves $1.17 \times$ speedup against CGBN-1 on A100 GPU. As a highly optimized assembly code implementation, SOS can minimize number of IMAD instructions and reduce stall caused by register bank conflict and register move. \oursys{} reduce can deliver an additional $1.39 \times$ performance, since we use as few additions and subtractions as possible to replace the IMAD operations in the reduce phase. 

\subsection{Application Performance}
\label{subsec:eval_app}
In the final set of experiments, we investigate the end-to-end effect of \oursys{} on the overall performance of \ecc applications. 
ECDSA is widely used in blockchain systems or blockchain databases for user identity authorization and verification for transaction security and integrity. For instance, in a blockchain database, each modification history of the database is recorded in the transaction. The ECDSA encryption safeguards the data from malicious tampering. %However, the inherently low throughput of the blockchain restricts the widespread application of blockchain-database.}
We evaluate the effect of \oursys{} to accelerate 
% the ECDH algorithm of PSI computation and
the ECDSA algorithm of a real-world permissioned blockchain system FISCO-BCOS~\cite{fisco2024}.

We run the blockchain on a four-node cluster, each with a 2.40GHz Intel Xeon CPU and 512 MB memory. The system transaction throughput without applying \oursys{} is approximately $5,948$ transactions per second, with the time breakdown revealing that the ECDSA signature generation and verification accounts for $37.2\%$ of the total time. Despite the significant performance improvement of our ECDSA, its throughput is constrained by other factors, such as consensus or other hash computations. After applying \oursys{}, the throughput can reach $9,313$ transactions per second, representing a performance improvement of $1.56 \times$.

%% file: conclusion.tex
\section{Conclusion}
ECC has a lower computational complexity and a smaller key size compared to \rsa{}, making it competitive for digital signatures, blockchain, and secure multi-party computation. Despite its superiority, ECC remains the bottleneck in these applications because the \ec{} operations in ECC are still time-consuming, which makes it imperative to optimize their performance. In this paper, we study how to optimize a batch of EC operations using the GAS mechanism and Montgomery’s Trick on GPUs. We propose locality-aware kernel fusion optimization and design multi-level cache management to minimize the memory access overhead incurred by frequent data access for point data and the intermediate results when batching EC operations. Finally, we optimize the operation performed most frequently $modmul$ in all types of EC operations.
% A significant obstacle in applying \ecc{} for these applications is the performance overhead of \ec{} operation. 
% This paper introduces \oursys{}, the general-purpose \ecc{} framework on GPU, designed for high-throughput \ec{} operations. 
Our results reveal that \oursys significantly improves the parallel execution efficiency of batch \ec{} operations and archives much higher throughput than the state of the art.